\documentclass{IEEEtran}
\ifCLASSINFOpdf
   \usepackage[pdftex]{graphicx}
\else
   \usepackage[dvips]{graphicx}
\fi

\usepackage[cmex10]{amsmath}
\usepackage{amsfonts}
\usepackage{algorithm}
\usepackage{algpseudocode}
\usepackage{subcaption}
\usepackage{amsmath}
\usepackage{ntheorem}
\theoremseparator{:}

\usepackage{booktabs} 
\usepackage{multirow}
\usepackage{multicol}
\usepackage{fnpct}
\ifCLASSOPTIONcompsoc \usepackage[caption=false,font=normalsize,labelfont=sf,textfont=sf]{subfig}
\else
\usepackage[caption=false,font=footnotesize]{subfig}
\usepackage[inkscapelatex=false]{svg}
\fi
\usepackage{tikz}
\makeatletter
\let\old@ps@headings\ps@headings
\let\old@ps@IEEEtitlepagestyle\ps@IEEEtitlepagestyle
\def\psccfooter#1{%

    \def\ps@IEEEtitlepagestyle{%
        \old@ps@IEEEtitlepagestyle%
        \def\@oddfoot{\strut\hfill#1\hfill\strut}%
        \def\@evenfoot{\strut\hfill#1\hfill\strut}%
    }%
    \ps@headings%
}
\makeatother
\usepackage[nolist]{acronym}
\usepackage[caption=false,font=footnotesize]{subfig}

\begin{acronym}
    \acro{ADN}{active distribution network}
    \acro{DER}{distributed energy resource}
    \acro{BGM}{balance group manager}
    \acro{BRP}{balance responsible party}
    \acro{TSO}{transmission system operator}
    \acro{DSO}{distribution system operator}
    \acro{GCP}{grid connection point}
    \acro{SOC}{second order cone}
    \acro{aFRR}{automatic frequency restoration reserve}
    \acro{PV}{photovoltaic}
    \acro{FCR}{primary frequency regulation}
    \acro{aFRR}{secondary frequency regulation}
    \acro{VPP}{virtual power plant}
    \acro{BESS}{battery energy storage system}
    \acro{OPF}{optimal power flow}
    \acro{SOE}{state of energy}
    \acro{MC}{Monte Carlo}
    \acro{AS}{ancillary services}
    \acro{JCC}{joint chance constraint}
    \acro{HP}{heat pump}
    \acro{GHI}{global horizontal irradiance}
\end{acronym}
\usepackage{titlesec}
\titlespacing{\section}{0pt}{1pt}{0.5pt}  
\titlespacing{\subsection}{0pt}{0.5pt}{0.5pt}
\titlespacing{\subsubsection}{0pt}{0.5pt}{0.5pt}
\usepackage{float}
\floatstyle{plaintop}
\restylefloat{table}

\begin{document}

\title{Probabilistic Flexibility Aggregation of \acp{DER} for Ancillary Services Provision}

\author{Matthieu Jacobs,~\IEEEmembership{Student Member,~IEEE}
        and Mario~Paolone,~\IEEEmembership{Fellow,~IEEE}
\thanks{The authors are with the Swiss Federal Institute of Technology of Lausanne, Switzerland, email: \{matthieu.jacobs,mario.paolone\}@epfl.ch. This work was sponsored by the Swiss Federal Office of Energy’s “SWEET” program and performed in the PATHFNDR consortium.}}

\maketitle

\begin{abstract}
This paper presents a grid-aware probabilistic approach to compute the aggregated flexibility at the \ac{GCP} of \acp{ADN} to allow the participation of \acp{DER} in \ac{AS} markets. Specifically, an \ac{OPF} method is used to compute the aggregated capability for the provision of multiple \ac{AS} at the \ac{GCP} of an \ac{ADN} over the full \ac{AS} provision horizon. The proposed method extends current approaches for the time-coupled flexibility aggregation, allowing to consider multiple services simultaneously, ensure cost-effectiveness of the used \acp{DER} and handle uncertainties in a probabilistic way. The disaggregation of the total flexibility into individual \acp{DER} power flexibilities accounts for the operational costs associated to the provision of different services. This ensures cost-effectiveness while maximizing the value of the aggregated flexibility, assuming known service prices. Empirical uncertainty sets are obtained to achieve a predefined coverage of the probability distribution in line with recent developments in the Nordic \ac{AS} markets. Finally, a feeder-decomposition approach is proposed to ensure applicability to realistic distribution networks with a large number of buses. Case studies show the effectiveness of the method, highlight the importance of network constraints and time-coupling and illustrate its applicability to realistic distribution systems.
\end{abstract}


\section{Introduction}
\acresetall
\label{sec:intro}
The large penetration of renewable energy generation and the increasing electrification of various processes change the paradigm of electricity system operation. Stochastic fluctuations of electricity generation, combined with concentrated peak demand, lead to more uncertainty in the operation of the electricity system. This is reflected in the increasing amounts of balancing services and the associated prices for \ac{aFRR} \cite{acer_market}. In response to these evolutions, the authors of \cite{entsoe_tsodso} stated that ''\textit{It is essential to take advantage of the opportunity to harness the valuable and increasing amount of resources at the distribution level for providing services for the overall benefit of the power system.}". New regulations allowing \acp{DER} to participate in energy and flexibility markets provide wholesale market benefits and new revenue streams for \acp{DER} owners and operators \cite{ieee_techreport_flex}. At the same time, \acp{DER}, aggregated in virtual power plants, provide a viable alternative to conventional generation for cost-effective provision of grid balancing services \cite{ieee_techreport_pe}. As a consequence, \acp{DER} are competing with conventional service providers, such as fuel-based generators, which are to be phased out in accordance with international greenhouse gas targets (e.g. \cite{acer_flex}). Additionally, most \acp{DER} are connected at the distribution level. This leads to increased loading of distribution systems and impacts the overall steady state and dynamic performance of the bulk power system \cite{ieee_techreport_bulk}. Stronger cooperation between \acp{DSO} and \acp{TSO} and appropriate strategies to make \ac{DER} flexibility available for the provision of grid services at the transmission level are required.

\subsection{\ac{DSO}-\ac{TSO} Interaction}
\label{sec:intro_cooperation}
This section briefly highlights the importance of \ac{DSO}-\ac{TSO} coordination, before Section \ref{sec:intro_agg} reviews existing approaches for the coordination mechanism of interest, flexibility aggregation. Readers interested specifically in coordination mechanisms are referred to \cite{ieee_techreport_flex}-\cite{coord_rev} and the references therein. The power systems community has largely acknowledged the need for closer cooperation between \acp{TSO} and \acp{DSO} \cite{ieee_techreport_bulk}. As the volume of services provided by \acp{DER} increases, \acp{DSO} must be actively involved to avoid issues at the distribution level \cite{coord_rev}. Furthermore, by only optimizing the operation at the distribution level, solutions that are locally optimal may not lead to a global optimum of the whole system. A broad overview of the main approaches for \ac{DSO}-\ac{TSO} coordination is provided in \cite{coord_rev}. The authors differentiate between three models based on the responsibilities of both system operators. In a first option, the \ac{DER} bids are directly transmitted to the \ac{TSO}, who selects and activates the required bids accounting for the \ac{DSO} constraints. The second approach considers that the \ac{DSO} first validates the bids based on its own operational constraints, before the \ac{TSO} selects the preferred bids. Finally, in the '\ac{DSO}-managed' model, the \ac{DSO} validates and aggregates the \ac{DER} bids, the \ac{TSO} selects the required aggregated flexibility and the \ac{DSO} sends the resulting activation commands. Assessing the aggregated flexibility in \acp{ADN} is beneficial for both the planning and operation \cite{ieee_techreport_flex}. 

\subsection{Flexibility aggregation approaches}
\label{sec:intro_agg}
We consider a \ac{DSO}-\ac{TSO} cooperation where the \ac{DSO} aggregates the \ac{DER} capabilities to offer flexibility to the \ac{TSO}. The relevant problem is to quantify and explicitly determine the available flexibility, while accounting for network and resource constraints and stochastic boundary conditions. \footnote{In this respect, the considered problem is different than approaches presented in \cite{dispatching_1} and \cite{dispatching_2}, where the operation of a distribution system under uncertainty is considered and works such as \cite{market}, where clearing of energy and flexibility markets is considered while accounting for \acp{DER}.} In this section, existing approaches for flexibility aggregation are discussed and the remaining issues are identified. A fundamental distinction between the methods reported in literature can be made based on the type of algorithm used \cite{sfoe_tsodso}. 
\par{i) \emph{Simulation-based Approaches}} \newline
The first group consists of approaches based on \ac{MC} simulations. Here, a significant number of scenarios are considered, each corresponding to a feasible actuation of the flexible resources. For each scenario, the corresponding power exchange at the \ac{GCP} is obtained. If all constraints are satisfied, the corresponding power exchange is labeled as a feasible point within the PQ plane. By performing a large number of simulations (with a consequent computation time), an accurate mapping of the power capability can be made. An example of such an approach is presented in \cite{mc_agg}.  Recent developments (e.g. \cite{MC_additional}) consider this problem and propose a Bayesian learning approach to alleviate the computational challenges. However, they still rely on an initial outer-box approximation of the flexibility set obtained through optimization.
\par{ii) \emph{Optimization Approaches: Exact Aggregation}}
On the other hand, optimization-based approaches can be divided in a group of methods that attempt to compute the exact power aggregation set at the \ac{GCP} and a group of methods that compute an approximate set, which can typically be described more easily. We further differentiate existing methods based on four important features of the flexibility aggregation sets: grid-awareness, time-coupling, cost-awareness and wether they account for uncertainty. As an example, \cite{exact_agg_wen} proposes an approach to compute the exact aggregated power flexibility set of multiple \acp{DER} without accounting for grid constraints. They also show that the number of constraints may be too large to solve the exact problem and have instead proposed a set of approximate models. Recognizing this issue, \cite{exact_agg_yi} proposes to cluster \acp{DER} based on the similarity of their capabilities and obtain an inner approximation through a specific type of polytope. The \acp{DER} in each cluster can then be exactly aggregated. Alternatively, in \cite{exact_agg_wen2024} a "geometric prototype" is constructed by selecting a subset of the equations forming the exact power aggregation set, keeping geometrical features of the exact model while reducing the computational complexity. Approaches attempting to identify the exact aggregated flexibility set, while simultaneously satisfying the grid constraints, also exist. In this case, a set of \ac{OPF} problems is typically solved. For instance, in \cite{exact_agg_silva} a set of non-approximated non-convex \ac{OPF} problems are solved to obtain points on the aggregated capability curve in the PQ plane until a convergence criterion based on the distance between the obtained points is satisfied. This approach also allows to include cost constraints. Similar approaches are also proposed in \cite{exact_agg_majmundar}, \cite{agg_multiOPF1}, \cite{agg_multiOPF2} and \cite{agg_multiOPFreview}. Explicit computation of flexibility costs for these approaches is proposed in \cite{agg_multiOPFcosts}, but costs are computed a posteriori for identified points of the flexibility map and consider a single time step. These approaches only allow to compute the aggregation set for a single time step as the solved \ac{OPF} problems cannot account for time coupling. To alleviate this problem, \cite{exact_agg_data_Li} proposes a network-informed data driven approach based on a classifier labeling samples to obtain an approximation allowing for temporal coupling. 
\par{iii) \emph{Optimization Approaches: Convex Approximations}} \newline 
While the above methods can obtain flexibility areas closer to the exact set, they are only applicable when neglecting grid constraints or considering a single time step or have to resort to approximations to make the problem tractable. Therefore, methods seeking convex inner approximations of the exact flexibility set are desirable. An approach for the estimation of a robust aggregation set is presented in \cite{flex_approx_robust}, where a linear power flow model is used. However, only single time steps are considered and \acp{DER} costs are not taken into account. In \cite{approx_agg_chen} a grid-aware approach, using a fixed-point linearization, is presented to compute the aggregate power flexibility over multiple time steps by approximating it as a hyperbox. However, uncertainty of stochastic prosumption is not accounted for. The authors attempt to model the costs, but only the one of the base trajectory, around which flexibility is offered, is included. In \cite{bernstein} the authors show that the flexibility aggregation set obtained through a hyperbox approximation is very conservative and that a larger one can be obtained with an inner ellipsoidal approximation. Using the same fixed-point linearization, the authors present tractable reformulations to obtain the maximum-volume ellipsoid under affine and quadratic disaggregation policies, while accounting for uncertainty through ellipsoidal uncertainty sets at each time step. An approximate approach decomposing the flexibility aggregation set in a "virtual battery" and "virtual generator" is presented in \cite{approx_agg_wang}. Uncertainty is accounted for through chance constraints assuming Gaussian probability distributions suitably transformed to deterministic constraints. The authors claim better results compared to the ellipsoidal inner approximation. The same authors extend this in \cite{approx_agg_wang2} with a piecewise fitting of the aggregated cost function for the different time steps. However, the cost function is computed by minimizing the operational cost for different points on the time-decoupled aggregation sets and fitting an affine function. Therefore, it does not account for the time-coupling of the flexibility and its influence on the costs. An alternative approach to the approximate aggregation problem is proposed in \cite{approx_agg_taheri}. By solving the disaggregation problem for samples selected in the flexibility aggregation space and labeling feasible points a convex ellipsoidal classifier is found. This ellipsoid is used as a surrogate solution space within which a polytope mapping the aggregated flexibility is determined. Uncertainty is accounted for when constructing the surrogate set. 
\subsection{Contributions of the Paper}
All the works above consider the flexibility aggregation problem to be the mapping of a single flexibility commodity at the \ac{ADN} \ac{GCP}. However, even though all flexibility services require power exchange, the provision of multiple ones (e.g. \ac{FCR} and \ac{aFRR}) have different implication on time-coupling and, therefore, the feasibility of an aggregation set. This is demonstrated in works regarding the optimal provision of \ac{AS} using \acp{BESS}, such as \cite{multiple_as_kazemi}. Furthermore, in the works considering uncertainty, forecast errors are either modeled robustly, leading to overly conservative solutions, or through chance constraints related to individual time steps. This does not accurately represent the impact of stochastic variables. Finally, although some works consider costs, none of these methodologies enforce cost-effectiveness of flexibility in an appropriately time-coupled manner. To summarize, none of the above approaches consider simultaneously network constraints, time-coupling and cost of the flexibility and the impact of uncertainty. The provision of multiple \ac{AS} is also not considered in state-of-the-art aggregation approaches. Therefore this paper makes the following contributions:
\begin{enumerate}
    \item We propose a multi-service flexibility aggregation approach based on (quadrant)-ellipsoidal sets.
    \item We integrate in the aggregation problem explicit constraints modeling the disaggregation and ensuring any selected flexibility is cost-effective.
    \item Uncertainty is accounted for in a probabilistic way through joint uncertainty sets, following the P90 requirement \cite{gade_p90} for the availability of power flexibility.
\end{enumerate}

\section{Problem Statement}
\label{sec:prob}
The problem considered in this work is the flexibility aggregation at the \ac{GCP} of \acp{DER} within \acp{ADN} with generic topologies (i.e. either meshed or radial) to provide multiple \ac{AS} at the transmission level. To accurately model the available flexibility, it is essential to consider both the capabilities of \acp{DER} and the constraints of the network. This ensures the aggregated flexibility does not adversely impact the operation of the distribution system. Additionally, the impact of uncertain prosumption and the time-coupling of the \acp{DER} flexibility capabilities is highly relevant. Consider the general formulation (\ref{eq:general}), where $p_0$ is the vector representing the power exchanged at the \ac{GCP} over all time steps, $\Omega_{p_0}$ is the set of feasible $p_0$ values, $p$ and $q$ are stacked vectors containing the vectors of controllable active and reactive power injections\footnote{Note that in this formulation, the controllable injections are assumed to be known, with a capability curve that is constrained by the uncertainties.} over all time steps and $\zeta \in \mathcal{U}$ is the stacked vector collecting the uncertainty drivers $\zeta_t$ over all time steps. 
\begin{subequations}
\label{eq:general}
\begin{align}
    \max_{\Omega_{p_0}, p, q} & \quad J(\Omega_{p_0}): \forall \zeta \in \mathcal{U}  \\
      \textrm{s.t.} \quad & G(p,q) + b(\zeta) = p_0 , \quad W(p,q) \leq z(\zeta)
    \end{align}
\end{subequations}
$J$ represents the flexibility maximizing objective, $G$ and $b$ represent the mapping between respectively the controllable resources and the uncertainty drivers and the slack power, while $W$ and $z$ represent the network and resource constraints. The goal of (\ref{eq:general}) is to determine the flexibility-maximizing set $\Omega_{p_0}$ of power exchanges at the \ac{GCP}. Any point within this set should be feasible for any realization of the uncertainty drivers. Directly obtaining such a set is in general untractable, even when neglecting grid constraints and using linear \ac{DER} models \cite{minkowski_summing}. The authors of \cite{bernstein} propose a convex, tractable approach to determine the maximum power flexibility of an \ac{ADN} by restricting the set of power exchange trajectories at the \ac{GCP} to an ellipsoid, following (\ref{eq:ellipsoid}) with $e$ the center and $E$ describing the axes and shape. Any point within this ellipsoid represents a vector of power exchanges at the \ac{ADN} for the considered horizon. The use of (\ref{eq:ellipsoid}) makes $\Omega_{p_0}$ explicit.
\begin{equation}
    \label{eq:ellipsoid}
    p_0 = E \xi + e, \quad ||\xi||_2 \leq 1
\end{equation}
In Section \ref{sec:prob_model} we summarize important results from \cite{bernstein} before we extend them in the next sections to determine the flexibility potential for the provision of multiple \ac{AS}. To this end, Section \ref{sec:prob_multiple} shows the aggregation of multiple \ac{AS}, Section \ref{sec:prob_pdf} proposes a probabilistic approach to integrate uncertainties, Section \ref{sec:prob_cost-effectiveness} introduces cost-effectiveness,  and Section \ref{sec:multifeeder} extends the method for multiple feeders.

\subsection{Network model and power aggregation}
\label{sec:prob_model}
The presented approach is applicable to any linear power flow model (e.g. as presented in \cite{jabr_pf}). In this work, the model introduced by \cite{bernstein_network} is used. Network constraints are linearized using a fixed-point equation, which can be interpreted as a linear interpolation of local power flow linearizations around two operating points, leading to a good global approximation. The linear grid model\footnote{An explicit expression for the current magnitudes is not provided in \cite{bernstein_network} as they can be written as second order cone constraints using the real and complex parts of the line currents. However, a linear model for the branch current magnitudes is obtained similarly as for the nodal voltage magnitudes.} is presented in (\ref{eq:grid_model}) with $p$ and $q$ representing the active and reactive controllable power injections. 
    \begin{equation} \label{eq:grid_model}
        p_0 = G\begin{bmatrix}
            p \\q
        \end{bmatrix} + b,   \; 
        |v| = K^v\begin{bmatrix}
            p \\q
        \end{bmatrix} + w, \;
        |i| = K^i \begin{bmatrix}
            p \\q
        \end{bmatrix} + d 
    \end{equation}
The coefficient matrix $G$ and constant coefficient $b$ represent an affine mapping for the power balance and $K^v$ and $K^i$ represent the linear coefficients mapping the power injections to the nodal voltage magnitudes $|v|$ and the branch current magnitudes $|i|$ respectively, with $w$ and $d$ the corresponding constant terms. Using the grid model (\ref{eq:grid_model}), the network constraints are written as (\ref{eq:network constraints}).
\begin{equation}
\label{eq:network constraints}
    v_{min} \leq |v| \leq v_{max}, 0 \leq |i| \leq I_{max}    
\end{equation}
By using linear approximations for all resource constraints, together with a linear power flow model and restricting the flexibility set to an ellipsoid, the flexibility-maximizing problem can be rewritten as (\ref{eq:original_linear}) \cite{bernstein} . 
\begin{subequations}
\label{eq:original_linear}
\begin{align}
    \max_{E, e, p} & \quad \textrm{log}(\textrm{det}(E)) \quad \textrm{s.t.} \quad  \forall ||\xi|| \leq 1, \forall \zeta \in \mathcal{U}   \\
      G & \begin{bmatrix}
            p \\q
        \end{bmatrix} + M_b\zeta + b_0 = E \xi + e, \; W \begin{bmatrix}
            p \\q
        \end{bmatrix} \leq M_z\zeta + z_0
\end{align}    
\end{subequations}
$M_b$ and $b_0$  and $M_z$ and $z_0$ are the affine mappings from $\zeta$ to $b$ and $\zeta$ to $z$ respectively. $W$ models the inequalities, including (\ref{eq:network constraints}). Equality constraints should be eliminated when possible in robust optimization problems \cite{RO}. To this end, the authors of \cite{bernstein} propose to exploit the structure of the problem. By writing $p = B_1x + B_2y$, with the columns of $B_1$ being an orthogonal basis for $G$ and the columns of $B_2$ spanning the null-space of $G$, equality constraints are eliminated. Finally, to make this problem tractable, an affine policy (\ref{eq:affine}) is introduced, with linear dependencies $K$ and $L$ on the position on the ellipsoid and the  uncertainties respectively and a constant $\gamma$. With these reformulations, the problem is given by (\ref{eq:original}). 
\begin{subequations}
\label{eq:original}
\begin{align}
     \max_{E,e,K,\gamma, L} & \quad \textrm{log}(\textrm{det}(E)) \quad  \textrm{s.t.} \quad   \forall ||\xi|| \leq1, \forall \zeta \in \mathcal{U} \\
       W_1 D^{-1} &(E\xi  + e - (M_b\zeta + b_0)) +  W_2 y  \leq   M_z\zeta + z_0 \quad  \label{eq:original_con}\\ \label{eq:affine}&   y = K\xi + \gamma + L \zeta \\
      &W_1 = WB_1, \; W_2=WB_2, \; D=GB_1
\end{align}    
\end{subequations}
To reformulate the robust constraints in (\ref{eq:original}) the worst case over the ellipsoidal aggregation set and uncertainty is considered for each constraint individually. Substituting (\ref{eq:affine}) in (\ref{eq:original_con}) and evaluating for the worst case uncertainties gives an upper bound for the uncertain part (\ref{eq:robust_unc}) denoted with the auxiliary variables $\alpha_i$ for each constraint $i$. Replacing this upper bound in (\ref{eq:original_con}) gives the deterministic constraint (\ref{eq:robust_con}). Together, this results in the deterministic problem (\ref{eq:robust}) where $\Theta$ represents the affine dependence on $\zeta$, $\nu$ collects the constant offsets and the subscript $i$ denotes the ith row.
\begin{subequations}
\label{eq:robust}
\begin{align}
    &\max_{E,e,K,\gamma, L}  \textrm{log}(\textrm{det}(E)) \quad \textrm{s.t.} \quad \forall i=1..m  \\ 
     & \alpha_i + w_{1,i} D^{-1}e  + w_{2,i} \gamma -\nu_i \leq 0  \label{eq:robust_con}\\
      & ||w_{1,i} D^{-1} E + w_{2,i} K|| + \max_{\zeta \in \mathcal{U}} ((w_{2,i} L - \theta_{i})\zeta) \leq \alpha_i \label{eq:robust_unc} 
\end{align}
\end{subequations}
\subsection{Resource Constraints}
\label{sec:prob_resources}
The power flexibility aggregated at the \ac{GCP} is provided by the controllable \acp{DER} in the \ac{ADN}. Although their character is typically truly non-linear \cite{variable_capcurves}, resource capabilities may be accurately described through linear constraints, as shown for example in \cite{mn_besscontrol}. This allows a linear representation of all the constraints limiting the flexibility provision, as required to eliminate the equality constraints in (\ref{eq:original_linear}). As set of representative \acp{DER}, we consider \acp{BESS}, \acp{HP} and \ac{PV} installations. The particularity of \acp{BESS} and \acp{HP} is that they introduce time-coupling, meaning the flexibility aggregation problem can not be solved for each $t$ separately. The battery active and reactive power $p_b,q_b$ and \ac{SOE} constraints $\forall t$ are:
\begin{subequations}
\begin{align}
    p_b^{min} \leq & p_b^t \leq p_b^{max}, \;  SOE^{min} \leq  SOE^t \leq SOE^{max}\\
    q_b^{min} \leq & q_b^t \leq q_b^{max}, \; SOE^{t+1} = SOE^t-p_b^t \Delta_t\
\end{align}    
\end{subequations}
The flexibility of \acp{HP} and \ac{PV} installations is slightly more complex to model as the uncertainty of the heat demand and the \ac{GHI} directly impacts the capability of these \acp{DER}. The following linear model for \acp{HP}, coupled to a temperature-dependent heat demand with a heat buffer tank, is used. The variables $T$ and $Q$ represent the temperature and heat respectively, while subscripts $bt$, $hp$, $demand$ respectively represent the buffer tank, heat pump and thermal demand. Finally, we define the buffer tank mass $m_{bt}$, the buffer tank losses $l_{bt}$, the heat pump model linear and constant terms $a_{hp}$ and $Q^0_{hp}$, the heat demand coefficient $h_{demand}$ and the comfort and environmental temperatures $T_c$ and $T_\infty$, related to the uncertainty vector by the affine mapping $H$.
\begin{subequations}
    \begin{align}
        T_{bt}^{t+1} - T_{bt}^t =& \frac{\Delta t}{4200m_{bt}} (Q_{hp}^t - Q_{demand}^t - l_{bt}) \\
        Q_{hp}^t = a_{hp}p_{hp}^t + & Q^0_{hp}, \; p_{hp}^{min} \leq p_{hp} \leq p_{hp}^{max} \\
        Q_{demand}^t = h_{demand}&(T_{c} - T^t_\infty),  \; T_{bt}^{min} \leq T_{bt}^t \leq T_{bt}^{max}
    \end{align}
\end{subequations}
The uncertainty appears in the environmental temperature $T_\infty=T_\infty^0 + H\zeta$. Similarly, the \ac{PV} installations are modelled as follows, with the subscript $MPP$ denoting the maximum power point power and $M_T$ the affine mapping between the power and the uncertainty vector.
\begin{subequations}
    \begin{align}
        0 \leq p_{PV}^t \leq p_{PV}^{MPP,t},  \; p_{PV}^{MPP,t} = p_{PV}^{MPP,t,0} + M_{PV}\zeta
    \end{align}
\end{subequations}

\subsection{Multiple services provision}
\label{sec:prob_multiple}
One of the main contributions of this work is to determine the flexibility capacity for multiple \acp{AS} simultaneously. \acp{DER} can provide multiple services, meaning the different services compete for the same power flexibility. An important feature of this formulation is that it allows to consider the aggregation of flexibility for different services in a non-hierarchical fashion. The multi-service problem (\ref{eq:multiservice}) is written as an extension of (\ref{eq:original_linear}). Note that the \ac{GCP} power for the different services $p_0^s$ is formulated here for active power services, which this work focuses on, as these are the services where time-coupling (e.g. due to limited energy storage) is most relevant. The formulation is however applicable for both active and reactive power services. Section \ref{sec:prob_cost-effectiveness} details how the objective is chosen to maximise the aggregation set, how \ac{DER} costs are accounted for and why they are not included in the objective function.
\begin{subequations}
\label{eq:multiservice}
\begin{align}
    \max_{p^s, \Omega_{p^s_0}} & \quad \sum_s J^s(\Omega_{p^s_0}) \quad \textrm{s.t.} \quad \forall \zeta \in \mathcal{U}, \forall s \in \mathcal{S} \\
    &\sum_s (W^s \begin{bmatrix}
            p^s \\q^s
        \end{bmatrix}) \leq M_z \zeta + z_0 \\
    & G^s \begin{bmatrix}
            p^s \\q^s
        \end{bmatrix} + M_b^s\zeta + b^s_0 = p^s_0, \quad p^s_0 \in  \Omega_{p^s_0}
\end{align}    
\end{subequations}

All services must be modelled separately as they might have different prices at the same time step, they have different power and energy requirements, they may have different time resolutions and activation requirements and their provision might cause different costs. For each service, the determined flexibility capacity at the \ac{GCP} represents the range of power set points that can be realized by the \ac{ADN}. Additionally, corresponding to standard market practices, the different services for which flexibility is offered, must be available independently. A robust representation of the exchanged power is therefore required. For linear grid and resource models this problem can be written in a linear way by using the superposition principle to sum the power contributions for the different services independently. A separate equality constraint can then be written for each service, linking the service slack power to the disaggregated contributions of the controllable resources. Geometrically, stacking different services can be seen as determining the maximum volume ellipsoids, weighted with the service prices, for which the Minkowski sum lies within the polytope representing the problem's constraints. Each service is related to the different constraints and the constraint satisfaction depends on all the services independently, reflecting independent selection and activation of the bids. This superposition also allows to specify additional constraints for individual services. Services with a different time resolution than the base time step can be forced to have the same values in blocks of time steps. \acp{DER} with specific characteristics can be excluded from providing certain services, by simply enforcing the power contribution for a specific service to be zero. Considering the different types of services, an important distinction must be made between symmetric (e.g. \ac{FCR}) and asymmetric services (e.g. \ac{aFRR}). Symmetric services, ranging from $-p^t_{0s}$ to $p^t_{0s}$ at each time step, can be naturally represented as ellipsoids, with some restrictions on the orientation. For asymmetric services, bids can be either for positive or negative power exchange, and any point between zero and the selected bid $p^t_{0s}$ must be feasible. A natural representation for these services, based on the ellipsoidal capability aggregation, comes in the form of quadrant ellipsoids. The representation of these (quadrant) ellipsoids is further detailed here below.
\begin{itemize}
    \item For symmetric services, the flexibility is represented as an ellipsoid that is restricted to be symmetric around all the axes. Without this restriction, the ellipsoids do not lead to valid \ac{FCR} ranges. Figure \ref{fig:ellipsoid} illustrates this condition. The red ellipsoid is not symmetric around the axes, meaning that the capacity for upward regulation does not necessarily match the capacity for downward regulation, as illustrated by the red dots. By restricting the ellipsoids to be symmetric around all axes, we obtain ellipsoids of the blue type, for which the blue dots illustrate the matching up/down regulation.
     With this restriction, $E^{sm}$ becomes diagonal and $e^{sm}$ vanishes, leading to (\ref{eq:symm_el}). Finally, without loss of generality, the elements of $E^{sm}$ are restricted to be positive as they represent the maximum symmetric power capacity that can be offered at any time step and both positive and negative power exchanges are included through the realization of $\xi$.
     \begin{subequations}
      \label{eq:symm_el}
    \begin{align}
        & p^{0,sm} = E^{sm} \xi^{sm}, \quad ||\xi^{sm}|| \leq 1 \\
        & \begin{bmatrix}
            p^{sm} \\q^{sm}
        \end{bmatrix} = (B_1^{sm} D^{{sm}^{-1}} E^{sm}+ B_2^{sm} K^{sm}) \xi \label{eq_controllable_sym}
     \end{align}         
     \end{subequations}
    \begin{figure}[h!]
    \centering
    \begin{subfigure}{0.34\linewidth}
    \includegraphics[width=1\linewidth]{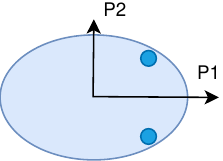}
    \caption{Symmetric Ellipsoid}
    \label{fig:symmell}
\end{subfigure}
\begin{subfigure}{0.34\linewidth}
    \includegraphics[width=\textwidth]{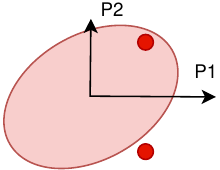}
    \caption{Asymmetric Ellipsoid}
    \label{fig:loadprofiles}
\end{subfigure}        
\caption{Ellipsoidal flexibility aggregation: 2D example.}
\label{fig:ellipsoid}
\end{figure}

    \item For asymmetric services, with upwards regulation, the flexibility is represented as a positive quadrant ellipsoid, restricted by (\ref{eq:asymmpos_el}). The restrictions imposed for the symmetric services are also enforced for the asymmetric services. This is needed to represent the services as quadrant ellipsoids, as off-diagonal terms may lead to negative power injections for a $\xi$ in the positive quadrant.
    \begin{subequations}
    \label{eq:asymmpos_el}
    \begin{align}
        & p^{0,ap} = E^{ap} \xi^{ap}, \quad ||\xi^{ap}|| \leq 1, \xi^{ap} \geq 0 \\
        & \begin{bmatrix}
            p^{ap} \\q^{ap}
        \end{bmatrix}  = (B_1^{ap} D^{{ap}^{-1}} E^{ap}+ B_2^{ap} K^{ap}) \xi \label{eq_controllable_pos}
    \end{align}        
    \end{subequations}
    \item For asymmetric services with downwards regulation, the flexibility set becomes a negative quadrant ellipsoid (\ref{eq:asymmneg_el}).
    \begin{subequations}
    \label{eq:asymmneg_el}
    \begin{align}
        & p^{0,an} = E^{an} \xi^{an}, \quad ||\xi^{an}|| \leq 1, \xi^{an} \leq 0\\
         & \begin{bmatrix}
            p^{an} \\q^{an}
        \end{bmatrix} = (B_1^{an} D^{{an}^{-1}} E^{an}+ B_2^{an} K^{an}) \xi \label{eq_controllable_neg}
    \end{align}
    \end{subequations}
\end{itemize}
Note that for all the services, both the slack power $p^{0,s}$ and the controllable injections $[p^s, q^s]^T$ are fully linear in $\xi$ with no constant component, as the service activation is uncertain and the flexibility set must include the zero activation option.

Next to the offered flexibility, the power exchanged at the \ac{GCP} that is not linked to any \ac{AS}, called baseload power, is also included. This work considers three use cases for the base power, referring to the different \ac{TSO}-\ac{DSO} coordination approaches. The base power can be uncontrollable, in which case it simply reflects the power flowing at the \ac{GCP} as a consequence of the uncontrollable prosumption. Alternatively, the base power can be controlled, for example by a centralized \ac{DSO} who optimizes the expected energy cost of operating the \ac{ADN}. In the following, this case is referred to as the 'baseload control'. Both these cases are occurrences of the third \ac{DSO}-\ac{TSO} coordination scheme, where the \ac{DSO} validates and transmits the aggregated bids to the \ac{TSO}. Finally, in the case of 'self-dispatching' the base power at different time steps is restricted to an predefined (ellipsoidal) set and controllable resources are activated to balance the uncertain prosumption and limit the uncertainty at the \ac{GCP}. This means the \ac{DSO} is responsible for its own imbalances and only the remaining flexibility can be offered to the \ac{TSO}. The modelling of baseload power for these cases is detailed here below.
\begin{itemize}
    \item In the self-dispatching case, the goal is to minimize the uncertainty on the baseload perceived at the \ac{GCP}. The baseload power is represented as an ellipsoid using the approach introduced Section \ref{sec:prob_model}. This naturally reduces the uncertainty on the base power exchange as any point within the ellipsoid must be feasible for all realizations of the uncertainty, meaning a larger ellipsoid would reserve more \acp{DER} flexibility. This leads to the following set of possible baseload powers, as in (\ref{eq:original}).
    \begin{align}
    \label{eq:self-dispatching_el}
    &p_0^{b} = E^b \xi + e^b, \quad ||\xi|| \leq 1 \\
    &p^b = (B_1^{b} D^{{b}^{-1}} (E^{b}\xi +(e^b-b(\zeta))) +  B_2^b (K^b \xi  + \gamma^b + L \zeta)   \notag      
    \end{align}
    \item For the controllable baseload case, two new variables $p_0^b$ and $y^b$ are introduced. These represent the expected slack power and controllable injections (\ref{eq:baseload_el}). In this case only the expected value $p^b_0$ is optimized. 
    \begin{equation}
    \label{eq:baseload_el}
     p^b = B_1^b D^{b^{-1}} (p_0^b - b(\zeta)) + B_2^b y^b 
    \end{equation}
    \item In the uncontrolled baseload case, the base slack power is the net result of the prosumption at the different nodes. This can be seen as a special case of the baseload control case, where $p^b=0$ and $y^b=0$
\end{itemize}

\subsection{Probabilistic Constraint Satisfaction}
\label{sec:prob_pdf}
The available power flexibility in \acp{ADN} depends on the stochastic prosumption, as it influences both the network constraints and the capabilities of \acp{DER}. Failing to account for these uncertainties leads to an overestimation of the \ac{ADN}'s power flexibility, while robustly ensuring the flexibility capacity is available can be overly conservative. Certain grid operators have already recognized this issue \cite{gade_p90}. To allow \acp{DER} with uncertain capabilities to participate in the \ac{AS} markets, the grid operator allows actors to offer flexibility services as long as the offered capacity is available at least 90\% of the time \cite{energinet} (i.e. according to the P90 requirement). To this end, the stochastic prosumption is handled in a probabilistic way in this work, with the availability requirement interpreted as a \ac{JCC} enforcing a 90\% probability that the the full flexibility is available at each time step. As \acp{JCC} are generally hard to solve, we determine a robust uncertainty set for the uncertainty drivers that guarantees a certain coverage $1-\epsilon$ of the empirical probability distribution. Note that uncertainty sets can be obtained for each time step separately, considering as stochastic variables $\zeta_t^u$ (e.g. with $||\zeta_t^u||\leq 1$ at time $t$ as was done in \cite{bernstein}). However, to guarantee a coverage of the joint probability distribution and jointly satisfy the problem constraints, a single uncertainty set for $\zeta = [\zeta^{u1},\zeta^{u1} ...\: \zeta^{u|\mathcal{U}|}]$ should be considered for all uncertainty drivers over the full horizon.  In this work, we assume a single uncertainty driver $\zeta^{u}$ suffices to model the stochastic fluctuations of all individual injections of a certain type (e.g. all PV injections are modelled with the same uncertain irradiance since, for distribution networks it is reasonable to assume they are located in the same geographical area.). Stochastic injections can be formulated as $p^u = \Bar{p}^{u} + M  \zeta^u \: \forall u \in \mathcal{U}$, with $M^u$ a linear mapping from the uncertainty drivers to the stochastic power injections\footnote{Consider for example PV plants. The power can be approximated as $P^{cap}*$\ac{GHI}, where the \ac{GHI} uncertain.} Ellipsoidal uncertainty sets are frequently used due to their simple parametric and numerical representation. Such uncertainty sets do not enforce simultaneous satisfaction of the worst case realization for all considered random inputs, thus often leading to less conservative solutions \cite{nemirovski}. Furthermore, stochastic uncertainty often naturally allows an ellipsoidal representation, (e.g. uncertainty drivers with Gaussian distributions allow exact ellipsoidal uncertainty sets \cite{joint_prob}.) A first approach to determine an ellipsoidal uncertainty set is to fit a multivariate Gaussian distribution to the uncertain prosumption data. Ellipsoidal uncertainty sets can also be constructed to ensure a $1-\epsilon$ coverage of the empirical probability distribution of the available data. A disadvantage of ellipsoidal sets is that they may inflate the uncertainty set in certain directions, especially when extreme values for the different variables are correlated, such as \ac{PV} at subsequent time steps. Therefore, we also consider a hyperbox uncertainty set covering the same proportion of the empirical samples. A comparison of these approaches is presented in Section \ref{sec:results_uncertainty}.

\subsection{Cost-Effective Flexibility Maximisation}
\label{sec:prob_cost-effectiveness}
Existing literature typically maximises the volume of the aggregation set when considering multiple time steps. However, when considering multiple \acp{AS}, the value of the services at each time step becomes relevant. Given that any $||\xi|| \leq 1$ can be selected, the realized benefit is not known a priori. Additionally, optimizing for the best case as an example would lead to the maximization of a norm, making the problem non-convex. Finally, as any point on the ellipsoid may be selected, the objective should reflect both the value of all points on the ellipsoid and the range within which the system operator can select the desired power reservation. To quantify this value, the different service prices are used. This reflects the willingness of system operators to pay and thus the service value for the system. Therefore the objective in (\ref{eq:obj}) is selected as a proxy to maximize the flexibility value. Here,the quantity of flexibility offered at each time step is represented by the diagonal elements of $E^s$. Denoting the service prices at time steps $t$ as $\pi^s_t$, the objective becomes: 
\begin{equation}
\label{eq:obj}
   J = \sum_s J^s \; =  \sum_s \sum_t \pi^s_t E^s(t,t)
\end{equation} 
In principle, the aggregation set may be modelled differently than through an ellipsoid. Two common approaches in the literature rely on the use of (i) polygonal aggregation or (ii) hyperboxes. However, in the case of a polygon, for which the parameters are unknown, the flexibility maximization problem becomes non-convex \cite{RO}. In the case of a hyperbox, the problem can be reformulated in a convex way but, for the considered objective, the aggregation set does not capture the tradeoff between different bids as a hyperbox enforces all bids to be satisfied simultaneously.

To realistically represent the flexibility \acp{ADN} can provide, it is crucial to take into account the operational costs for the provision of the different services. Accounting for these ensures the flexibility is cost-effective and thus "economically feasible" \cite{econ_feasibility}. Therefore, we derive a cost-effectiveness condition that can be formulated in a convex way under some mild assumptions on the cost functions of the \acp{DER}. The costs for providing the different services can be obtained through the affine disaggregation policy used to map the flexibility provided at the \ac{GCP} to the individual \acp{DER}. The controllable injections can be obtained from (\ref{eq_controllable_sym}), (\ref{eq_controllable_pos}) and (\ref{eq_controllable_neg}).
\begin{equation}
    p^s = (B_1^s D^{s^{-1}} E^s + B_2^s K^s) \xi
\end{equation}
A sufficient condition for the convexity of the problem is that the costs for the \acp{DER} must be linear with respect to the allocated power capacity. This is a reasonable assumption as can easily be seen for the example of \acp{BESS} with equivalent cycles. For linear cost functions, the flexibility cost becomes:
\begin{equation}
    C^s = c^{s^T} |(B_1^s D^{s{-1}} E^s + B_2^s K^s) \xi|
\end{equation}
where $c^s$ is a cost vector containing the linear cost coefficients for the \acp{DER} for the provision of service $s$. This expression can be further simplified, as the flexibility capabilities are represented as (quadrant) ellipsoids, which means we can force all controllable injections to have the same sign. This means that they all contribute to the flexibility. 
\begin{equation}
\label{eq:positivity}
    B_1^s D^s{^{-1}} E^s + B_2^s K^s \geq 0
\end{equation}
The costs and benefits can then be equivalently reformulated as in (\ref{eq:costeffectiveness}) for service ellipsoids with positive elements. 
\begin{equation}
    \label{eq:costeffectiveness}
    C^s = c^{s^T} (B_1^s D^{s{-1}} E^s +B_2^s K^s)|\xi|,  \quad
    B^s = g_s^T E^s |\xi|  
\end{equation} 
Constraining the benefits to be higher than the costs ensures cost-effectiveness (\ref{eq:costeffectiveness_con}) for any set of activated services.
\begin{equation}
g^{s^T} E^s - c^{s^T} (B_1^s D^{s{-1}} E^s +B_2^s K^s) \geq 0 \label{eq:costeffectiveness_con}
\end{equation}
One might desire to directly optimize the net benefit (\ref{eq:costeffectiveness_con}). However, two key issues appear. First, the net benefit depends on the selected flexibility meaning the real benefit is expressed as:
\begin{equation}
\label{eq:net_benef}
    B^s_{net} = \sum_s (g^{s^T} E^s - c^{s^T} (B_1^s D^{s{-1}} E^s +B_2^s K^s) |\xi|^s
\end{equation}

Finally, to avoid biasing the energy available in the baseload case and thus unfairly increasing the profit by emptying storage assets, two constraints are added. 
\begin{align}
    \sum_t (p^{b}_0 - b) = 0 \label{eq:baseload_balance}\\
    \sum_t (c^{0,t}p^{b}_{0,t} - \sum_rc^{r^T}_t |p^{b,t}|) \geq 0  \label{eq:baseload_cost}
\end{align}
The first one (\ref{eq:baseload_balance}) enforces the integral of the difference of baseload power with respect to the expected net load to be zero. The second (\ref{eq:baseload_cost}) ensures the power balancing accounts for operational costs, where $r$ indexes the different \acp{DER}.
Including these cost-effectiveness considerations, the final problem is given by Problem (\ref{eq:full}). The impact of this controlled baseload on the problem constraints is modelled through (\ref{eq:baseload}) and the influence of the uncontrollable prosumption and corresponding adjustment of the base power is included in (\ref{eq:prosumption}). Note that the subscript $u$ refers to the number of uncertainty sets to ensure the formulation remains generic. The influence of all the services and the selected baseload control is combined in (\ref{eq:summing}) and (\ref{eq:ineq}) where $e^0$ and $\gamma$ disappear for the baseload case. We introduce auxiliary variables $\alpha^i$ for each constraint and $\alpha^i_s$ to constrain the impact of the different services on each constraint. 
Reformulating the problem with robust constraints and restricting the aggregated flexibility set to (quadrant) ellipsoids for all services yields the multiservice aggregation problem (\ref{eq:full}).
\begin{subequations}
\label{eq:full}
\begin{align}
    & \max_{E^s, K^s}  \quad \sum_s \sum_t \pi^s_t E^s(t,t) \\
      \textrm{s.t.} \; & ||w^{sm}_{1,i} D^{sm^{-1}} E^{sm} + w^{sm}_{2,i} K^{sm}|| \leq\alpha_i^{sm} \label{eq:symm}\\
         & w^{ap}_{1,i} D^{ap^{-1}} E^{ap} + w^{ap}_{2,i} K^{ap} \leq \epsilon_i^{ap} \label{eq:asymmpos}\\
          & ||\epsilon_i^{ap} || \leq \alpha^i_{ap}, \quad \epsilon_i^{ap} \geq 0 \label{eq:asymmpos2}\\
    & w_{1,i^{an}} D^{an^{-1}} E^{an} + w^{an}_{2,i} K^{an} \geq \epsilon_i^{an} \label{eq:asymmneg} \\
          &  ||\epsilon_i^{an} || \leq \alpha_i^{an}, \quad   \epsilon_i^{an} \leq 0  \label{eq:asymmneg2}\\
        &  (w^0_{2,i} L^{0} - \theta_{u,i}) \zeta \leq \lambda_i \quad \forall \zeta \in \mathcal{U}\label{eq:prosumption} \\
    & \alpha^i \geq \sum_s \alpha_{s,i} + \sum_t \lambda_{t,i} \label{eq:summing} \\
      & \alpha_i + \beta_i + w^0_{1,i} D^{-1}e^0  + w^0_{2,i} \gamma^0 -\nu_i \leq 0 \label{eq:ineq} \\
    & g^{s^T} E^s - c^{s^T} (B_1^s D^{s{-1}} E^s +B_2^s K^s) \geq 0 \label{eq:costeffectiveness_only} \\
    & B_1^s D^s{^{-1}} E^s + B_2^s K^s \geq 0 \label{eq:positivity_full} \\
    &\textrm{if self-dispatching}: \notag\\
    &||w^0_{1,i} D^{0{-1}} E^0 + w^0_{2,i} K^0||  \leq \alpha_i^0 \label{eq:self-dispatching}  \\ 
       & \textrm{if baseload:}  \notag\\
       & w^0_{1,i} D^{0^{-1}} p_0 + w^0_{2,i} y^0 \leq \alpha^0_i \label{eq:baseload} \quad  \\ &\sum_t (p^{b}_0 - b) = 0 \label{eq:baseload_balance_full}\\
    &\sum_t (c^0_t p^{b}_{0,t} - \sum_rc^{r^T}_t |p^{b,t}|) \geq 0  \label{eq:baseload_cost_full} 
\end{align}    
\end{subequations}
Finally, depending on the representation of the uncertainties (\ref{eq:prosumption}), the effect of uncertainty is reformulated in (\ref{eq:unc_ellipsoid}) and (\ref{eq:unc_hb_1}) and (\ref{eq:unc_hb_1}) for ellipsoid and hyperbox uncertainty sets respectively.
\begin{subequations} \label{eq:uncertainty_sets}
    \begin{align}
        &\lambda \geq ||w^0_{2,i} L^{0} - \theta_{i}|| \label{eq:unc_ellipsoid} \\
        &\lambda \geq \sum_u(w^0_{2,i,u} L^{0}_u - \theta_{u,i}) \zeta_u^{\max} \label{eq:unc_hb_1}\\
        &\lambda \geq \sum_u(w^0_{2,i,u} L^{0}_u - \theta_{u,i}) \zeta_u^{\min} \label{eq:unc_hb_2}
    \end{align}
\end{subequations}

\subsection{Multifeeder Aggregation}
\label{sec:multifeeder}
For scalability, an additional step allowing to aggregate multiple feeders connected to the same primary substation is developed. In this case, the aggregated flexibility is first computed for each feeder separately, by solving (\ref{eq:full}). In a second step, using the obtained aggregation sets, a further aggregation set for all the considered feeders is obtained by solving a problem similar to the single feeder aggregation. The contributions of the different feeders replace the contributions of the \acp{DER} and the resource capabilities are replaced by the feeder ellipsoids $E_f^s$ for each service $s$, leading to \ac{SOC} constraints. As all \acp{DER} constraints are embedded in the feeder ellipsoids, only the transformer rating constraints need to be added in this multifeeder aggregation problem. Given that any set of flexibility bids must be feasible irrespective of the total baseload power, the transformer rating is adjusted based on the range of possible baseload powers computed in the feeder aggregation problem. For consistency, the same objective is used as in the single feeder problem, requiring the positivity constraint (\ref{eq:positivity}) to model cost-effectiveness. This leads to the following formulation, where (\ref{eq:multifeeder_ellipsoids}) reflects the feeder flexibilities and (\ref{eq:multifeeder_transfo}) ensures the transformer, with upper and lower power limits $p^{max}_{transfo}$ and $p^{min}_{transfo}$, is not overloaded.
\begin{subequations}
\begin{align}
    & \max_{E^s, K^s} \quad \sum_s \sum_t \pi^s_t E^s(t,t)  \\ 
    \textrm{s.t.} \quad  &||E_f^{s^{-1}}(B_1^s D^{s^{-1}}E^s + B_2^s K^s)\xi|| \leq 1, ||\xi|| \leq 1  \label{eq:multifeeder_ellipsoids}\\
    &\sum_s (W_1^s D^{s^{-1}} E^s + W_2^s K^s) \xi \leq b , ||\xi|| \leq 1 \label{eq:multifeeder_transfo}\\
    &b = [p^{max}_{transfo} - p^{max}_{base}, p^{min}_{transfo} + p^{min}_{base}]^T 
\end{align}    
\end{subequations}
Since all feeders are fed by the same primary substation transformer, the individual aggregation problems can be solved separately by enforcing a robust slack voltage constraint at each feeder. Indeed, by allowing only solutions that satisfy all feeder constraints for voltages in between $1-\delta$ and $1+\delta$ pu, the problems can be decoupled while maintaining overall feasibility. This is done by duplicating the voltage and current network constraints for the extreme values of the considered slack voltage range. The total flexibility to be provided by the multifeeder \ac{ADN} is then disaggregated first between the different feeders, according to the multifeeder disaggregation policy and then further disaggregated within each feeder, according to their individual policies.

\section{Results}
\label{sec:results}
\begin{figure}[h!]
    \centering
\begin{subfigure}{0.65\linewidth}
    \includegraphics[width=\textwidth]{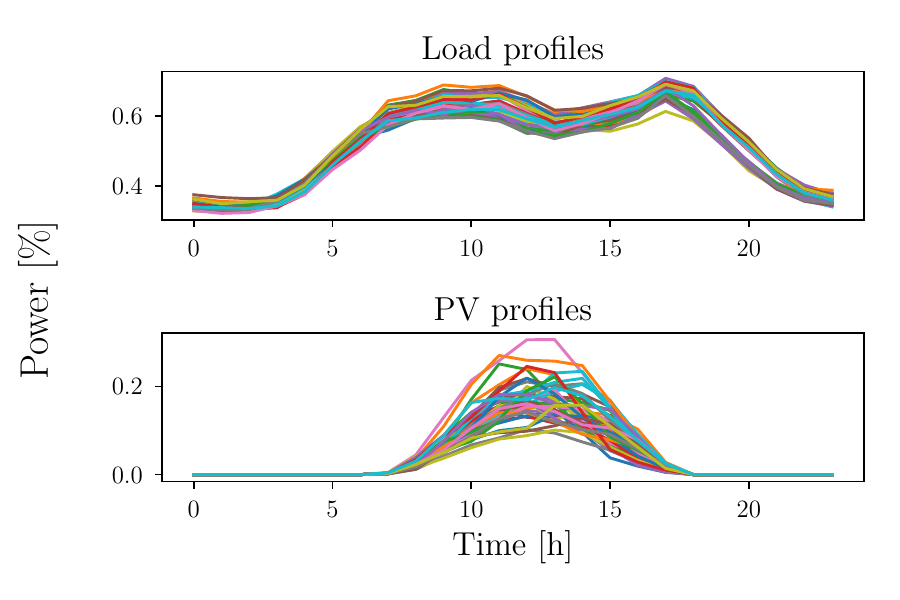}
    \caption{Load (top) and PV scenarios (bottom).}
    \label{fig:loadprofiles}
\end{subfigure}  \\
    \begin{subfigure}{0.65\linewidth}
    \includegraphics[width=1\linewidth]{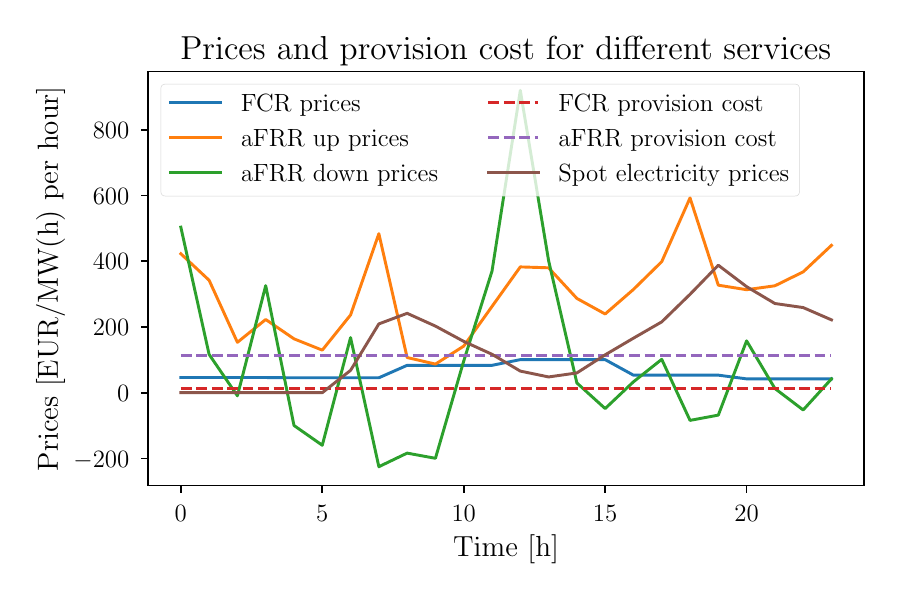}
    \caption{Prices for service provision and operation costs.}
    \label{fig:costs}
\end{subfigure}
\caption{Prosumption and costs data, from \cite{RTE}, \cite{dwd} and \cite{load_data}.}
\label{fig:inputdata}
\end{figure}

This section demonstrates the approach for a number of test cases. First results for a the IEEE33 benchmark system are shown when considering only \acp{BESS} for the baseload and self-dispatching cases. Then the approach is extended to multiple feeders, interconnected to a common primary substation, showing its effectiveness when aggregating the flexibility of realistic distribution systems. Next, the problem is extended to different \acp{DER}, showing how they participate in the aggregation. Finally the impact of network constraints is discussed. Common inputs for the service prices and the uncertainty sets are used. Input data is obtained from the Réseau de Transports d'Electricité \cite{RTE} (for the prices), Deutscher Wetterdienst \cite{dwd} (for the PV production) and historical residential data from \cite{load_data} for the loads. Figure \ref{fig:inputdata} shows the inputs used for the case studies. 
\begin{table}[htbp]
    \centering
    \caption{System Component Parameters}
    \label{tab:system_params}
    \resizebox{0.9\linewidth}{!}{%
    \begin{tabular}{@{}lcccccccccccccccccccccc@{}}
        \toprule
        & \multicolumn{8}{c}{\textbf{PV}} 
        & \multicolumn{4}{c}{\textbf{BESS}} \\
        \cmidrule(lr){2-9}\cmidrule(lr){10-13}
        \textbf{Node} 
            & 5 & 11 & 14 & 18 & 20 & 25 & 28 & 32 & 8   & 17   & 25  & 33   
            \\
        \textbf{Rating [kW]} 
            & 100 & 200 & 100 & 500 & 100 & 50 & 200 & 150 & 1000 & 4000 & 2000 & 1000 
             \\
        \bottomrule
    \end{tabular}}

    \vspace{0.5em}

    \resizebox{0.8\linewidth}{!}{%
    \begin{tabular}{@{}lccccccccccccccccccccccc@{}}
        \toprule
        & \multicolumn{11}{c}{\textbf{Load}} \\
        \cmidrule(lr){2-12}
        \textbf{Node}        & 2 & 3 & 4 & 6 & 8 & 9 & 10 & 11 & 13 & 15& 16 \\
        \textbf{Rating [kW]} & 90 & 120 & 60 & 200 & 60 & 60 & 45 & 60 & 120 & 60& 60 \\
        \hline
        \textbf{Node} & 17 & 19 & 21 & 22 & 23 & 24 & 26 & 27 & 29 & 30 & 31\\
         \textbf{Rating [kW]} & 90 & 90 & 90 & 90 & 420 & 420 & 60 & 60 & 200 & 150 & 60 \\
        \bottomrule
    \end{tabular}}
    \caption{Power Rating of \acp{DER} in the IEEE33 network.}
\end{table}
The results presented below all consider as flexibility services the provision of \ac{FCR} and \ac{aFRR} (up and down) as examples of symmetric and asymmetric services. To construct the uncertainty sets $1200$ from the $40x40=1600$ scenarios are used (i.e. combining the load scenarios with the weather ones) and $400$ are kept for a posteriori simulation of the flexibility provision.
\begin{figure}
    \centering
    \includegraphics[width=0.7\linewidth]{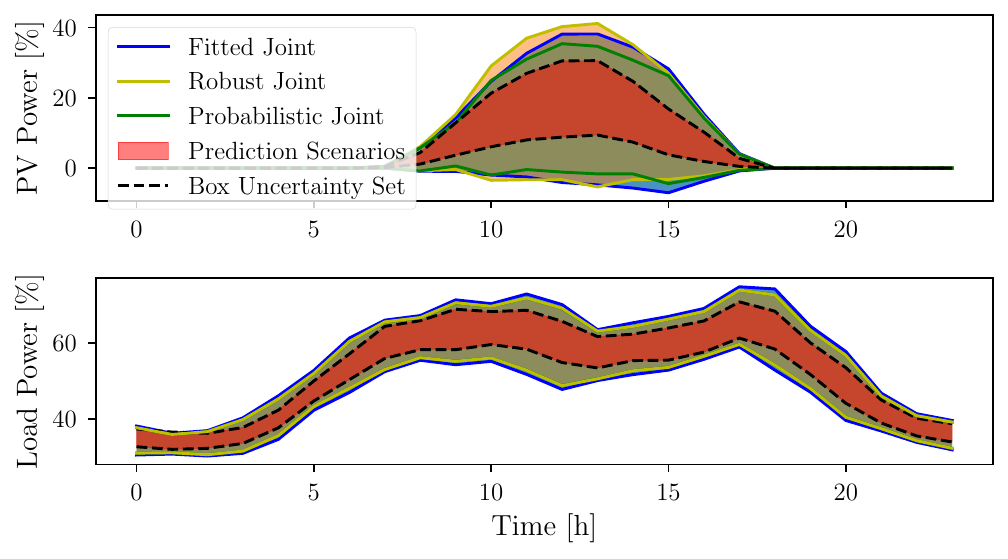}
    \caption{Uncertainty sets for the different ellipsoidal approaches and the hyperbox approach.}
    \label{fig:ellipsoids}
\end{figure}

\subsection{Comparison of the Uncertainty Sets}
\label{sec:results_uncertainty}

\begin{figure} [h]
\centering
  \includegraphics[width=0.90\linewidth]{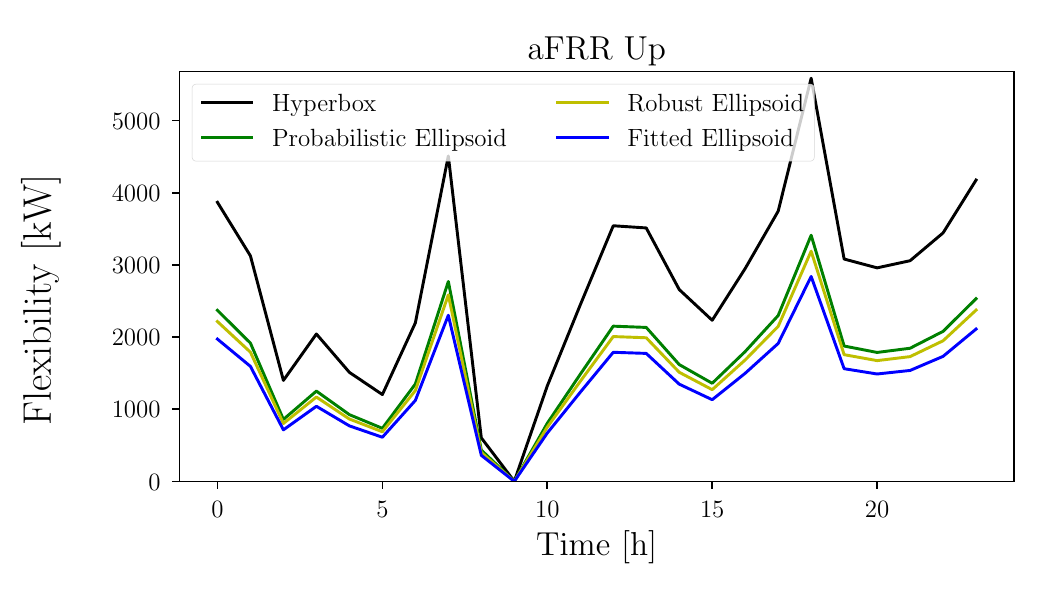}%
  \caption{Impact of uncertainty set on aggregated flexibility.}
  \label{fig:agg_unc_comp}
\end{figure}

This section shows the impact of the uncertainty representation on the obtained aggregated flexibility. Ellipsoidal and hyperbox uncertainty sets are obtained by ensuring a proportion $1-\epsilon$ (here set to $0.9$) of the in-sample scenarios are within the uncertainty sets. One approach to obtain such a set is based on fitting a multivariate Gaussian distribution to the prediction scenarios. As the load and PV uncertainties are not truly normally distributed, this leads to an overly conservative uncertainty set. Alternatively, robust ellipsoidal sets can be obtained by determining the minimum volume ellipsoid containing all points in prediction set \cite{boyd}. Assuming this set is representative for the future realizations allows to obtain probabilistic uncertainty sets containing a proportion $1-\epsilon$ of all the points. For example, by computing the mutual distance between all points and selecting the one with the lowest maximum distance to the $1-\epsilon$ points with the smallest distance to this point. Figure \ref{fig:ellipsoids} shows the boundaries of the ellipsoidal sets obtained in different ways. Representing the ellipsoids is not possible due to the large number of dimensions. Instead, the figure shows the range of possible values at all time steps for the different approaches, (i.e. the extreme points in all dimensions). The bounds for a robust ellipsoid considering all the in-sample scenarios are shown in yellow, the probabilistic uncertainty set assuming a Gaussian distribution is shown in blue and the set ensuring $90\%$ coverage of the in-sample scenarios is shown in green. The range of prediction scenarios and historical realizations used to obtain the uncertainty sets are also shown. Finally, the figure also shows the bounds of the hyperbox uncertainty set, which corresponds to the maximum and minimum values of the in-sample scenarios. Figure \ref{fig:agg_unc_comp} shows the aggregated flexibility for all cases (only \ac{aFRR} up is selected by the optimization problem in this case). Clearly, the hyperbox uncertainty set is the least conservative. The constraint violation rates in Table \ref{tab:unc_viol} confirm this trend, with all ellipsoidal sets leading to much lower violation rates than the allowed $10 \% $.

\begin{table}[h]
    \centering
    \resizebox{0.99\linewidth}{!}{
    \begin{tabular}{|c|c|c|c|c|}
    \hline
        Aggregation Set & Hyperbox & 90\% ellipsoid & Robust Ellipsoid & Gaussian Ellipsoid\\ \hline
        Violation Rate [\%]& 6.25 & 0 & 0 & 0.75 \\\hline
    \end{tabular}}
    \caption{Constraint Violation Rates}
    \label{tab:unc_viol}
\end{table}

\subsection{Benchmark Distribution System}
\label{sec:results_basic}
\begin{figure}[h!]
    \centering
    \begin{subfigure}{0.78\linewidth}
    \includegraphics[width=1\linewidth]{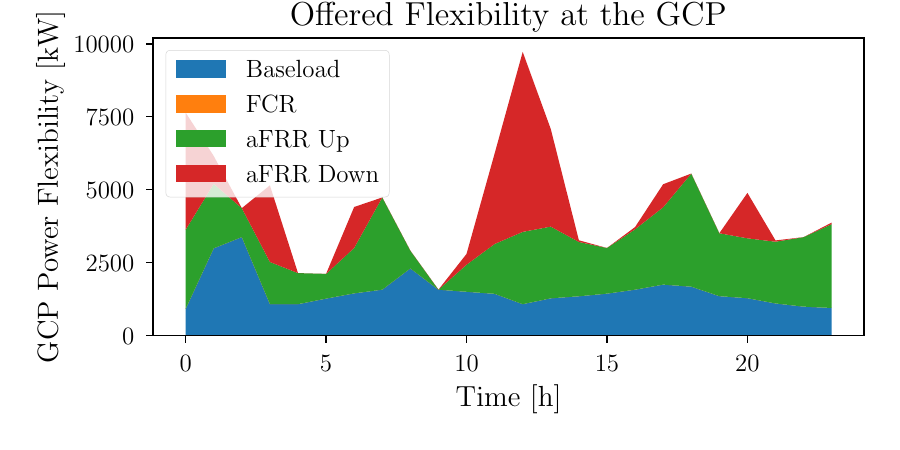}
    \caption{Aggregated flexibility offered at the feeder \ac{GCP}.}
    \label{fig:results_IEEE33_baseload_GCP}
\end{subfigure}
\\ 
\begin{subfigure}{0.78\linewidth}
    \includegraphics[width=\textwidth]{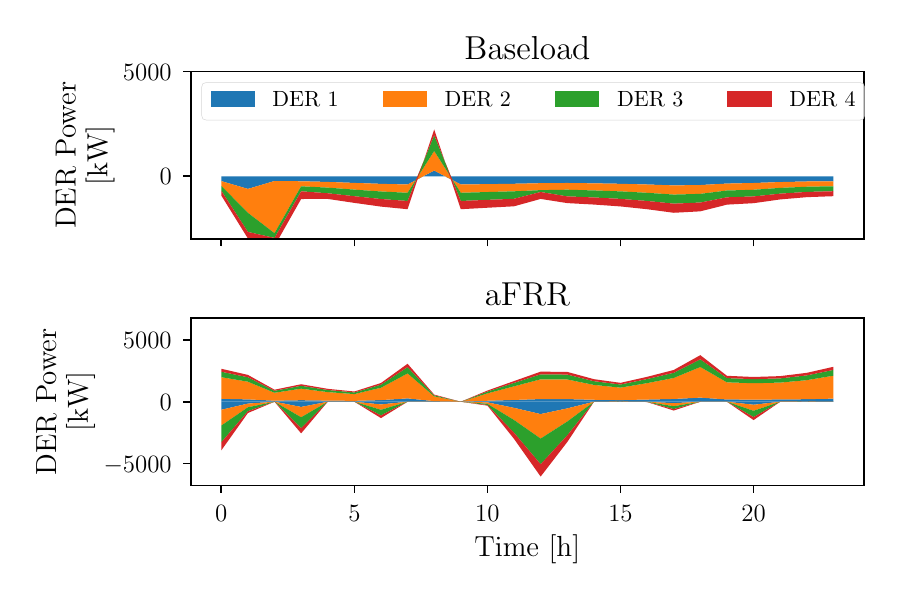}
    \caption{Contribution of the resources to the offered flexibility.}
    \label{fig:results_IEEE33_baseload_resources}
\end{subfigure}        
\caption{\raggedright Aggregated flexibility for the IEEE33 (baseload).}
\label{fig:results_IEEE33_baseload}
\end{figure}

\begin{figure}[h!]
    \centering
    \begin{subfigure}{0.78\linewidth}
    \includegraphics[width=1\linewidth]{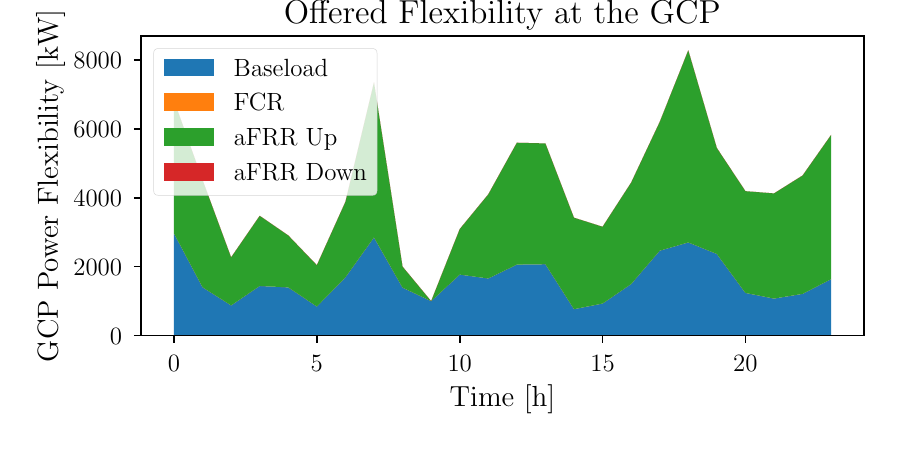}
    \caption{Aggregated flexibility offered at the feeder \ac{GCP}.}
    \label{fig:results_IEEE33_self_GCP}
\end{subfigure}
\\
\begin{subfigure}{0.78\linewidth}
    \includegraphics[width=\textwidth]{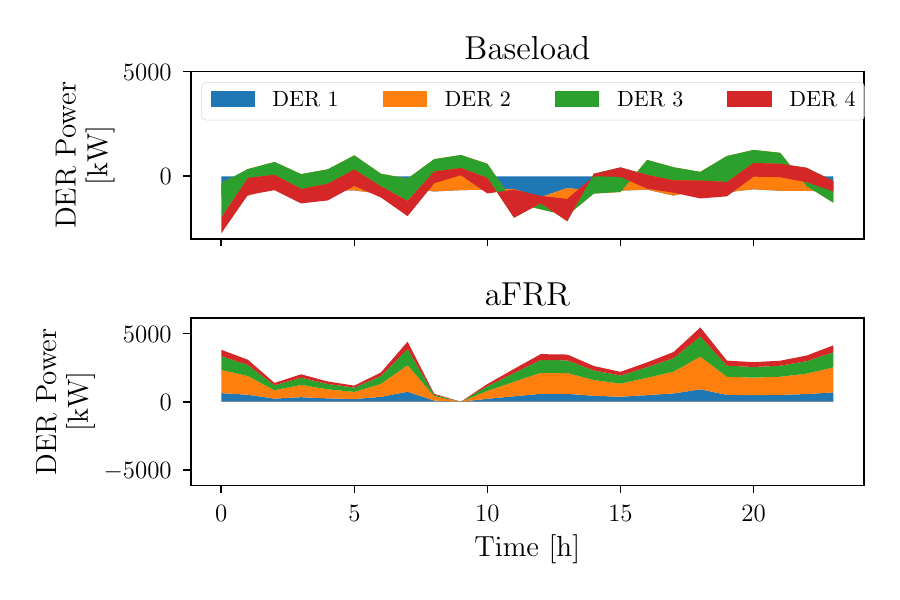}
    \caption{Contribution of the resources to the offered flexibility.}
    \label{fig:esults_IEEE33_self_resources}
\end{subfigure}        
\caption{Aggregated flexibility for the IEEE33 (selfdispatching).}
\label{fig:results_IEEE33_self}
\end{figure}

The IEEE33 benchmark feeder containing $1.3MWp$ of PV injections and $2.6MWp$ load is selected as a benchmark system. Four controllable \acp{DER} are considered, in this case all \acp{BESS} with a total energy capacity of $24MWh$ and a total power capacity of $8MW$. The \ac{BESS} capacities are obtained by increasing the storage capacity until the self-dispatching version of the problem became feasible, allowing a comparison between the different aggregation versions. Time-coupled uncertainty sets are used for both PV and load. Specifically, the hyperbox uncertainty sets obtained by considering the empirical $90\%$ coverage are used. Figure \ref{fig:results_IEEE33_baseload_GCP} shows the results when only the base power trajectory is controlled in a stacked plot. In this case more flexibility can be offered at the transmission level as uncertainties only impact potential congestions and other grid constraints but do not reserve energy or power capacity of the flexible \acp{DER} at the distribution level. Only \ac{aFRR} is selected due to the higher service prices in this case, with no \ac{FCR} provision for both the baseload and self-dispatching due to the lower prices.
Figure \ref{fig:results_IEEE33_self} shows the results for the self-dispatching case, in which the flexibility is first used to compensate local uncertainties before services are offered to the upper-level system. Only upward \ac{aFRR} is offered due to the limited flexibility. Note that the \ac{GCP} baseload power shown in Figure \ref{fig:results_IEEE33_self_GCP} does not represent the total local balancing energy budget as both upwards and downwards capacity are reserved. Much less flexibility is available in the self-dispatching case, compared to the baseload case, as a large energy budget is reserved to balance deviations of the stochastic prosumption from the baseload. To show the utilization of the \acp{DER} storage assets and validate the aggregation sets, we run \ac{MC} simulations, where for the point in the aggregation set that maximizes the provided flexibility, we sample the load and weather conditions from the out-of-sample scenario set. Out of the 400 scenarios, 25 are outside of the considered uncertainty set ($6.25\%$), leading to 25 violations for the self-dispatching case. For the baseload case, these scenarios outside of the uncertainty set do not lead to violations. This occurs because the actuation of controllable \acp{DER} is more dependent on the uncertainty for the self-dispatching case and the resource constraints are the most stringent in this use case. Other quantities such as the \ac{ADN}'s nodal voltages and branch currents are validated later in Section \ref{sec:results_grid}. 

\subsection{Flexibility Potential in Realistic Distribution Systems}
\label{sec:results_realsystem}
\begin{figure}[h!]
    \centering
    \begin{subfigure}{0.75\linewidth}
    \includegraphics[width=1\linewidth]{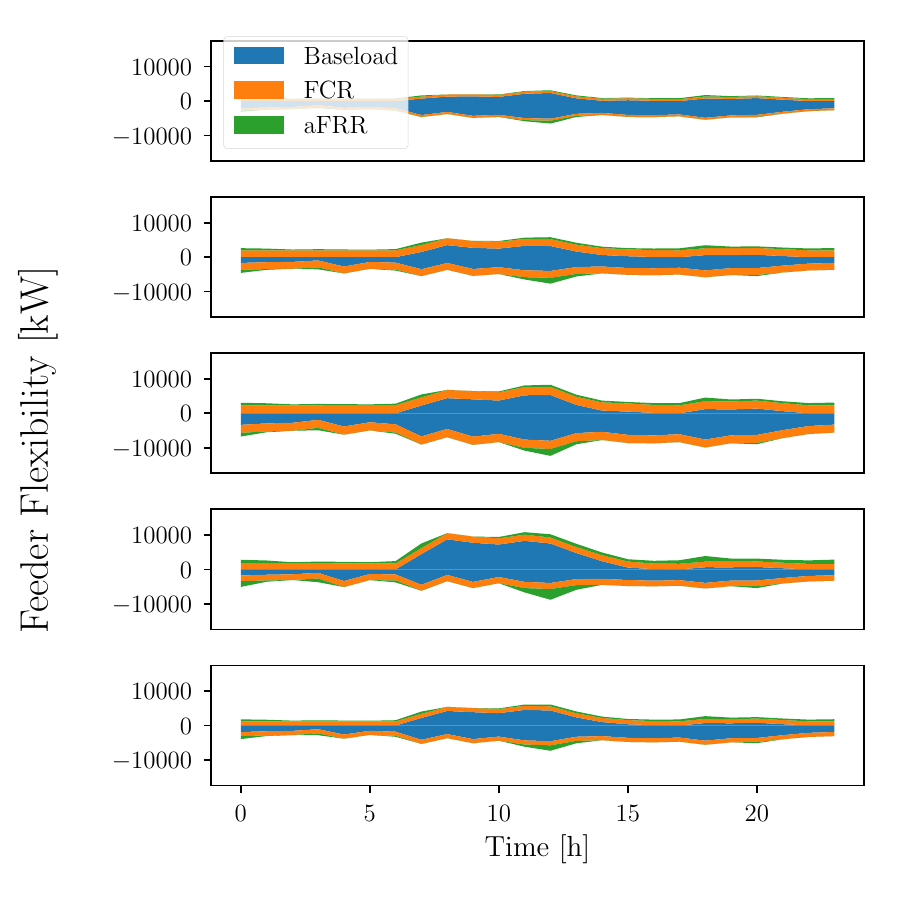}
    \caption{Aggregated flexibility offered at each feeder \ac{GCP}.}
    \label{fig:results_DSO_feeders}
\end{subfigure}
\\ 
\begin{subfigure}{0.75\linewidth}
    \includegraphics[width=\textwidth]{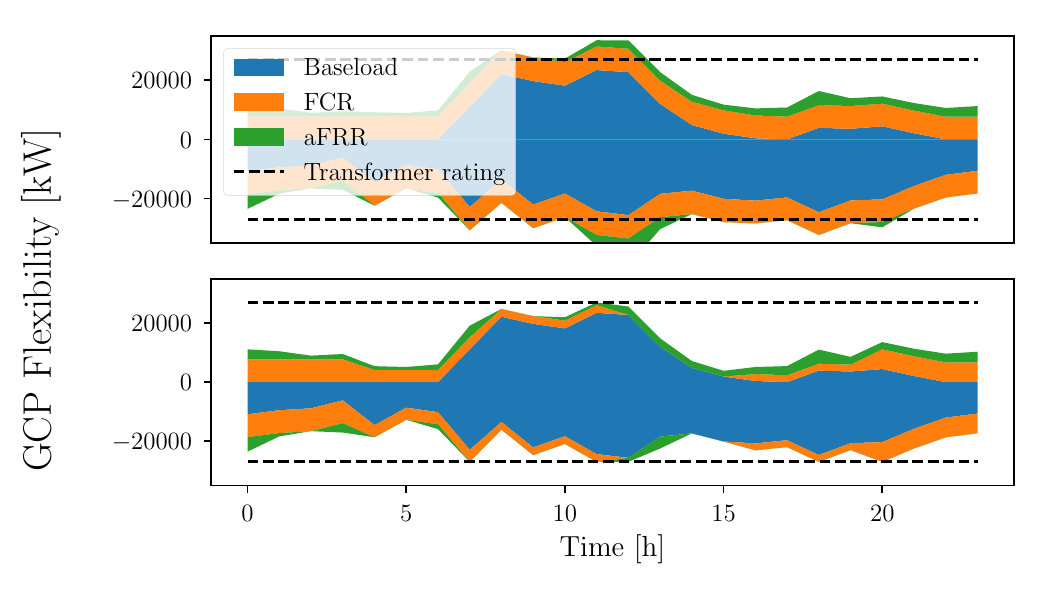}
    \caption{\raggedright Aggregated \ac{ADN} flexibility for two transformer ratings (top: $30MW$, bottom: $27MW$). The dotted line represents the tightened transformer rating on both.}
    \label{fig:results_DSO_agg}
\end{subfigure}        
\caption{\raggedright Flexibility for a multifeeder \ac{ADN} (baseload).}
\label{fig:DSOagg}
\end{figure}

 To demonstrate the multifeeder approach presented in Section \ref{sec:multifeeder}, we solve the aggregation problem for five similar \acp{ADN} obtained from the synthetic networks presented in \cite{gupta_syntheticnetworks}, with the network IDs given in Table \ref{tab:multifeeder}. The loading and PV integration for all networks are added in Table \ref{tab:multifeeder} together with the controllable \acp{DER} characteristics. The considered voltage range at the slack node of the individual \acp{ADN} or feeders is set to $\delta = 0.02 \: pu$. Figure \ref{fig:DSOagg} shows the results for two cases of the transformer rating at the full network \ac{GCP}. The service price of \ac{FCR} was considered double the value reported in Figure \ref{fig:costs} to show the provision of all services. Reducing the transformer rating to $27MW$ limits the available flexibility. For the symmetric \ac{FCR} service, this leads to a reduction in both directions although the transformer constraint is only binding for additional power consumption.
\begin{table}[]
    \centering
    \resizebox{0.80\linewidth}{!}{
    \begin{tabular}{|c|c|c|c|c|c|}
    \hline
        Network IDs from \cite{gupta_syntheticnetworks} & ID 1 & ID 2 & ID 4 & ID 7 &ID 8 \\ \hline
        Number of nodes & 23 & 19 & 27 & 24 & 17 \\\hline
        Load Rating [kW] & 3205& 2531& 5035  & 2589 & 2911 \\\hline
        PV Capacity [kW] &1380& 3085 & 4496& 9236&4456 \\\hline
        \ac{BESS} power capacity [kW] & 1800&4400& 5200& 6600& 3600\\\hline
        BESS energy capacity [kWh]&3600&8800& 10400& 13200& 7200\\\hline
        Number of loads & 18&10& 17& 8& 9\\\hline
        Number of PV injections &5&9 & 10& 16& 8\\\hline
        Number of \acp{BESS} & 3&3& 4 &4 &2 \\ \hline
    \end{tabular}}
    \caption{Characteristics of \acp{ADN} for the multifeeder case.}
    \label{tab:multifeeder}
\end{table}
For the application of the proposed aggregation approach, the computational complexity of the method is also relevant. Due to the complex nature of the problem, the computation time increases unacceptably when directly considering large distribution networks. The computational complexity of the problem is mostly due to the large number of conic constraints, required to reformulate the robust problem. The number of conic constraints in this problem can be expressed in (\ref{eq:cones}) as a function of the key problem dimensions. The first one is the number of timesteps $T$. The other problem parameters affecting the complexity are the numer of considered services $n_s$, the number of uncertainty sets $n_u$, the number of nodes $N_n$ and lines $N_l$, the number of controllable \acp{DER} and storage assets $N_s$.
\begin{equation}
\label{eq:cones}
    n_{cones} = (n_{s}+n_{u}) * (N_{n}+N_{l}+2*N_c+N_{s})*2*T
\end{equation}
The multifeeder approach allows to decompose large distribution networks per feeder to keep the problem tractable. For the two cases on the IEEE33 benchmark case, the computations times were 30 and 18 minutes for the self-dispatching and the baseload case respectively. For the multifeeder aggregation, the mean and max time per feeder were 24 and 45 minutes respectively. We recall that thanks to the robust voltage constraints, each feeder problem can be solved in parallel. The problem combining the feeders takes a few seconds. Solving the multifeeder problem in one go (without the disaggregation in feeders) was intractable on the adopted computer (a MacBook M1 Max with 32GB RAM).

\subsection{Extension to multiple \acp{DER} types}
\begin{figure}[h!]
    \centering
    \begin{subfigure}{0.78\linewidth}
    \includegraphics[width=1\linewidth]{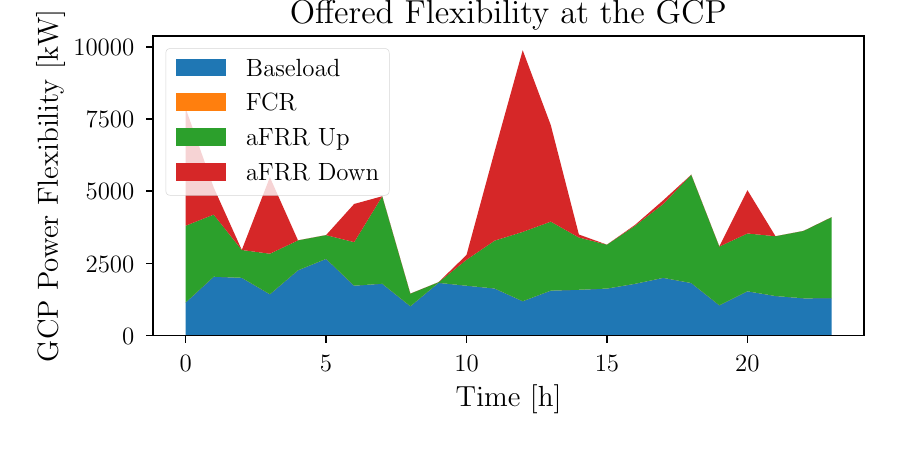}
    \caption{Aggregated flexibility offered at the feeder \ac{GCP}.}
    \label{fig:results_multiDER_GCP}
\end{subfigure}
\\ 
\begin{subfigure}{0.78\linewidth}
    \includegraphics[width=\textwidth]{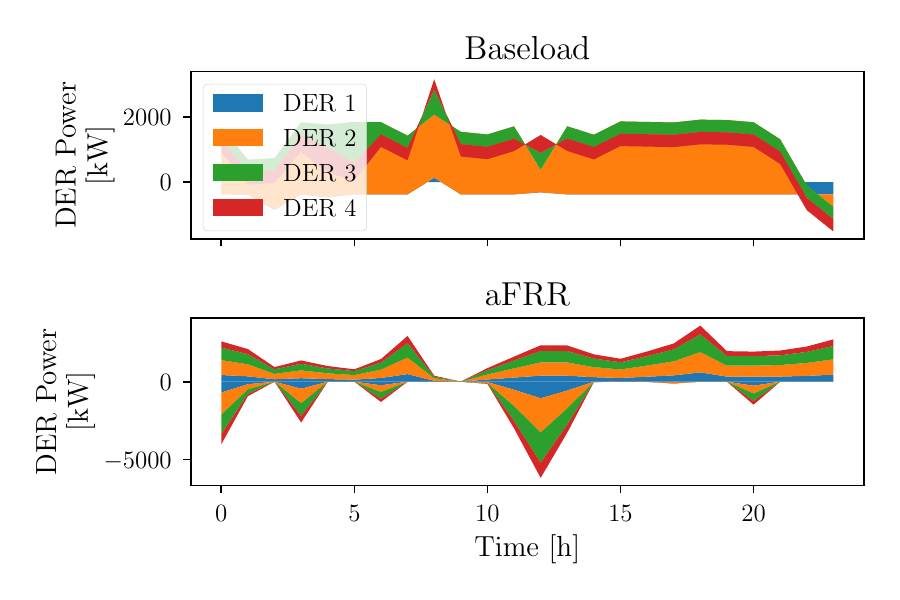}
    \caption{Contribution of \acp{BESS} to the offered flexibility.}
    \label{fig:results_multiDER_resources}
\end{subfigure}
\\ 
\begin{subfigure}{0.78\linewidth}
    \includegraphics[width=\textwidth]{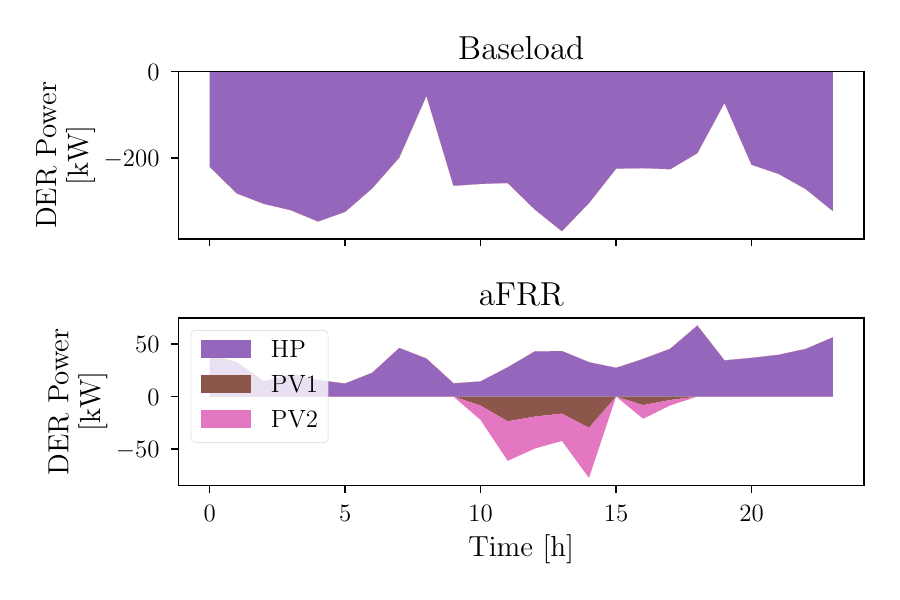}
    \caption{Contribution of other resources to the offered flexibility.}
    \label{fig:results_multiDER_resources}
\end{subfigure}
\caption{\raggedright Aggregated flexibility for the combined \ac{DER} case.}
\label{fig:results_multiDER}
\end{figure}

\begin{figure}[h!]
    \centering
    \begin{subfigure}{0.7\linewidth}
        \centering
        \includegraphics[width=0.95\linewidth]{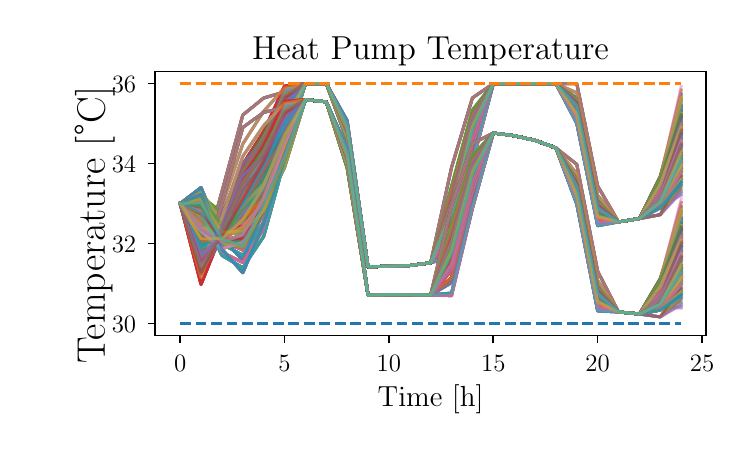}
        \caption{\ac{HP} temperature for the out-of-sample scenarios.}
        \label{fig:results_SOEhp_der}
    \end{subfigure}
    \\
    \begin{subfigure}{0.7\linewidth}
        \centering
        \includegraphics[width=0.95\linewidth]{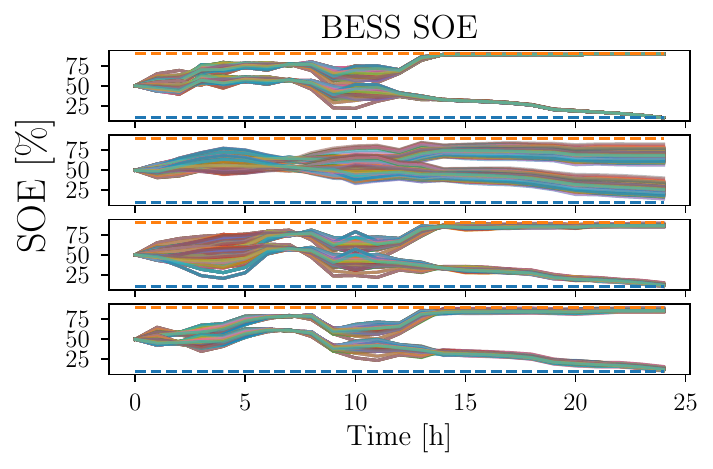}
        \caption{\acp{BESS} SOE for the out-of-sample scenarios.}
        \label{fig:results_SOE_der}
    \end{subfigure}
    \caption{Evolution of the storage for the combined \ac{DER} case.}
    \label{fig:DER_SOE}
\end{figure}

In this section, we show the impact of allowing \ac{PV} installations and \acp{HP} to participate in the flexibility provision. The two first \ac{PV} installations in the IEEE33 case, with power ratings increased to $500MWp$ and $800MWp$, are considered controllable, together with a \ac{HP} with a maximum electric power consumption of $250kW$. Figure \ref{fig:results_multiDER} shows which services are offered and   how the different types of \acp{DER} participate in the flexibility provision. Consistent with the operation of the different \acp{DER}, the \acp{BESS} are allowed to provide all services, while the \acp{HP} do not participate in the \ac{FCR} provision and the \ac{PV} installations only participate in the \ac{aFRR} provision and do not participate to the compensation of prosumption uncertainty. 
Figure \ref{fig:DER_SOE} shows the evolution of the \ac{SOE} of the \acp{BESS} and the temperature evolution for the \ac{HP} buffer tanks for the 400 out-of-sample scenarios used to validate the selected flexibility aggregation set. This clearly shows that both the \ac{HP} and \ac{PV} installations participate in the flexibility provision. The \ac{PV} installations only provide \ac{aFRR} down, as curtailment is disallowed to avoid participation in the compensating of the prosumption uncertainty. The \acp{HP} participate both to the baseload control and the \ac{aFRR} provision. This also allows the \acp{BESS} to increase their \ac{aFRR} provision. The out-of-sample analysis led to constraint violations in $2.75\%$ of the cases, with $11.25\%$ of the cases outside of the uncertainty set.

\subsection{Importance of grid-aware flexibility aggregation.}
\label{sec:results_grid}
\begin{figure}[h]
    \centering
    \begin{subfigure}{0.7\linewidth}
\includegraphics[width=1\textwidth]{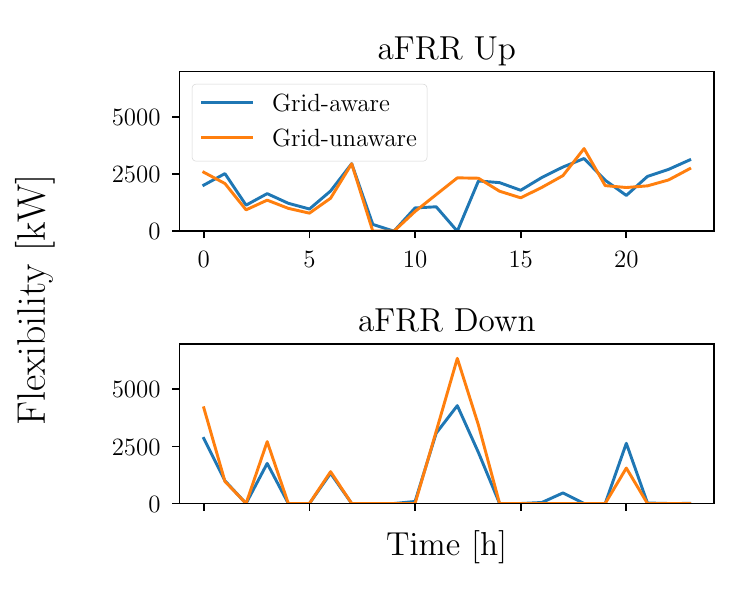}
    \caption{\raggedright Flexibility with and without network constraints.}
    \label{fig:grid_agg}
\end{subfigure} \\
\begin{subfigure}{0.7\linewidth}
    \centering
    \includegraphics[width=1\textwidth]{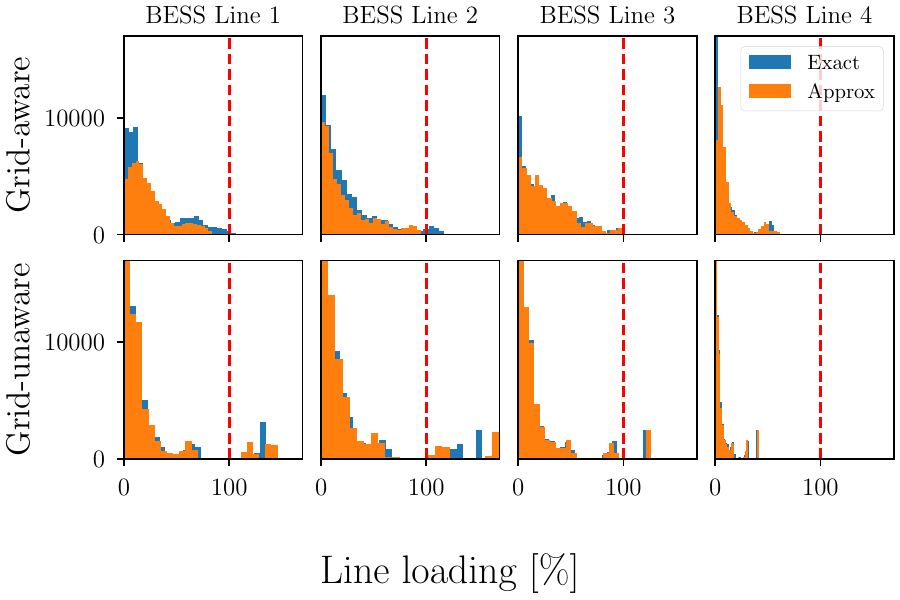}
    \caption{Currents for the grid-aware and grid-unaware cases.}
    \label{fig:results_grid_currents}
    \end{subfigure} 
    \caption{\raggedright Impact of network constraints.}
\end{figure}

This section shows how the network constraints impact the aggregated flexibility. To this end, the aggregated flexibility for the IEEE33 network is compared with the case where all network constraints are neglected. To create additional congestions highlighting the importance of modeling the network constraints, we set all line ampacity limits to a third of their nominal value. For the baseload case, removing network constraints means uncertainty plays no role when considering only \acp{BESS} as flexible \acp{DER} as none of the constraints are affected by uncertainty drivers. Therefore the full flexibility can then be used for \ac{TSO} requirements. Figure \ref{fig:grid_agg} shows the aggregated flexibility obtained with and without grid constraints. When the grid constraints are not accounted for, more flexibility can be offered at time steps with higher forecast service prices (cfr Figure \ref{fig:costs}). Due to the time-coupling, more flexibility is offered in the grid-aware approach for some time steps. Note that the additional flexibility offered in the grid-unaware case is the result of an overestimation of the available flexibility and is not deliverable due to the violation of grid constraints. Also here, no \ac{FCR} is provided. Figure \ref{fig:results_grid_currents} shows histograms of the line loading for the lines connecting the \acp{BESS} for the grid-aware and grid-unaware cases. Currents are computed a posteriori for a set of \ac{MC} simulations where both the flexibility exchanged and the uncertain prosumption are varied. The frequent violations of the line constraints in the grid-unaware case show the importance of accounting for grid constraints when determining the available flexibility. The results also show that the exact currents slightly differ from the ones predicted by the linear model in the optimization. Therefore in some cases, the true currents may exceed the ampacity limits, as shown for example in Figure \ref{fig:results_grid_currents} for \ac{BESS}2, but in general the limits hold.

\section{Conclusions}
\label{sec:conclusion}
In this work, we develop a new method to compute the aggregated flexibility of \acp{DER} hosted in \acp{ADN} for the provision of multiple \ac{AS}. The proposed method accounts for the uncertainty of prosumption in a probabilistic way through ellipsoidal uncertainty sets and ensures cost-effectiveness by accounting for \acp{DER} operation costs. The value of flexibility from a system operators point of view is maximized, assuming the service prices are known. Finally, we demonstrate the applicability of the method to realistic distribution grids by adding robust voltage constraints and combining the aggregated flexibility of multiple feeders while accounting for the \ac{ADN} transformer rating. The proposed framework improves the coordination between \acp{DSO} and \acp{TSO} by representing the available flexibility for different ancillary services separately. The presented approach represent the capability of \acp{DER} at the distribution level to provide flexibility at the transmission level and thus increase the number of market participants. The affine disaggregation policies for each \ac{AS} further allow to directly control the \acp{DER} in accordance with the selected flexibility. Additionally, by integrating the costs of the different resources, the aggregated flexibility can be represented using \ac{SOC} constraints, which could be directly integrated in conic markets, such as advocated by \cite{conic_markets}. Further work will include investigating approaches to integrate uncertainties without the need for pre-computed uncertainty sets.

\bibliographystyle{IEEEtran}
\bibliography{main.bib}
\appendix
\subsection{\acp{BESS} Operational Costs}
\label{sec:appendix_bess}
\begin{figure}[h]
    \centering
    \includegraphics[width=0.65\linewidth]{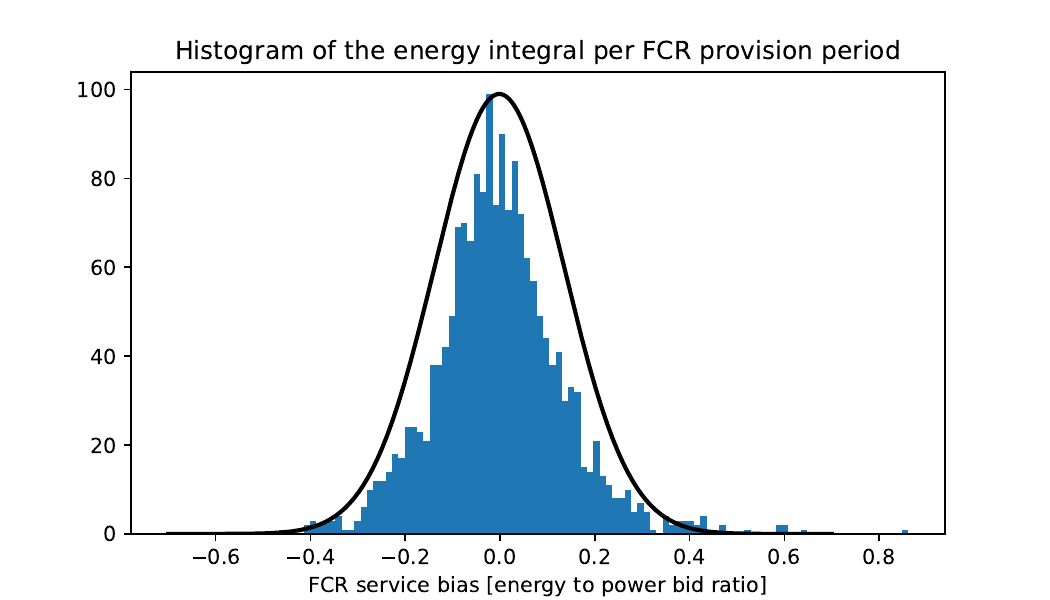} \\
    \includegraphics[width=0.65\linewidth]{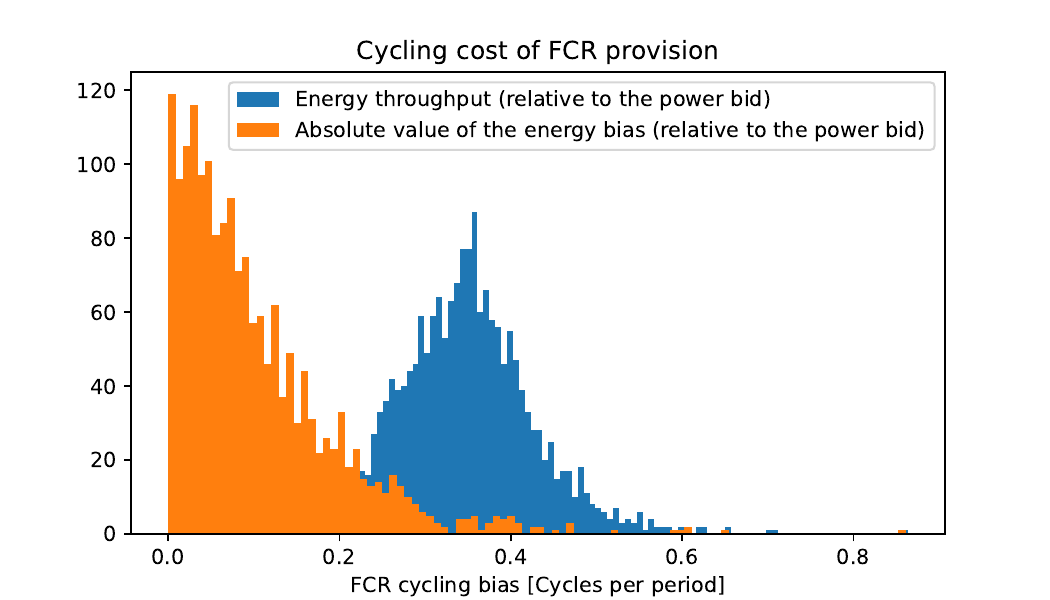}
    \caption{Equivalent \ac{BESS} cycles and \ac{SOE} needs for \ac{FCR}.}
    \label{fig:bess_cycles}
\end{figure}

    For \acp{BESS} the operational costs for providing the chosen flexibility services can be expressed based on the equivalent cycles required to adjust the power exchanges $n^{op}_{cycles}$. Equivalent cycles relate the energy throughput to the cycling aging and assign an operational cost to the battery. Using equivalent cycles, the operational cost $c^{op}$ can be expressed as a function of the investment costs $c^{inv}$ and the rated number of equivalent cycles $N_{cycles}$, through the cost per cycle $c^{cycle}$. 
\begin{equation}
    c^{op} = n^{op}_{cycles}c^{cycle}, \quad 
    c^{cycle} = c^{inv}/N_{cycles}
\end{equation}
Based on historical data for the frequency in continental Europe \cite{RTE} the required energy storage and number of equivalent cycles for the provision of FCR can be obtained and normalized by the power bid. Figure \ref{fig:bess_cycles} shows histograms of the energy bias per 4-hour period relative to the power bid when offering FCR (assuming the bid is equal to the battery capacity). The energy bias is defined as the change in \ac{SOE} over an \ac{FCR} provision period and determines how much energy capacity should be reserved to provide the \ac{FCR} service. The bottom figure also shows the energy throughput in a \ac{BESS}, obtained as the integral of the absolute value of power over the FCR provision periods. Based on this, we find that for more than $95\%$ of the cases, the relative energy bias is smaller than approximately $25\%$ and the relative energy throughput is smaller than $50\%$ when considering 4-hourly bidding periods. Therefore, we reserve an energy capacity equal to $25\%$ of the offered power flexibility for \ac{FCR}. For \ac{aFRR}, commands are sent directly by the \ac{TSO}. Following the merit order of balancing energy offers, some may be fully activated, while others are not used. In this case, the relative energy bias and throughput are set to one to ensure cost-effectiveness irrespective of the activation level and maintain a linear cost function.

\end{document}